\documentclass[preprint2]{aastex}

\newcommand{\gton}            {\objectname[\[BE83\] IR 309.92+00.48]{G309.92+0.48}}

\newcommand{\gtfo}            {\objectname[\[BE83\] IR 351.16+00.70]{G351.16+0.70}}

\newcommand{\gf}              {\objectname[GAL 005.89-00.39]{G5.89-0.39}}

\newcommand{\goo}             {\objectname[GAL 011.11-00.40]{G11.11-0.40}}

\newcommand{\goe}             {\objectname[GPSR5 18.147-0.284]{G18.15-0.28}}

\newcommand{\gso}             {\objectname[GAL 061.48+00.09]{G61.48+0.09}}

\newcommand{\gsvo}            {\objectname[GAL 070.29+01.60]{G70.29+1.60}}

\newcommand{\gsvsv}           {\objectname[GPSR 77.965-0.007]{G77.96-0.01}}

\newcommand{\kms}             {$\:\mathrm{\:km~s^{-1}}$}
\newcommand{\mac}             {$\:\mathrm{mag/arcsec^2}$}

\newcommand{\methanol}        {CH$_3$OH}
\newcommand{\halphasvs}       {H76$\alpha$}
\newcommand{\halphaef}        {H85$\alpha$}
\newcommand{\halphaoon}       {H109$\alpha$}
\newcommand{\halphaooo}       {H100$\alpha$}
\newcommand{\halphaonono}     {H110$\alpha$}
\newcommand{\water}           {H$_2$O}

\newcommand{\dgal}            {D$_{\mathrm{gal}}$}
\newcommand{\dsun}            {D$_\sun$}

\newcommand{\ks}              {$K_s$}

\newcommand{\hks}             {$H - K_s$}
\newcommand{\jh}              {$J - H$}
\newcommand{\uchii}           {UC~H~{\sc ii}}
\newcommand{\hii}             {H~{\sc ii}}
\newcommand{\avi}             {A$_\mathrm{v}$}
\newcommand{\nlyc}            {N$_\mathrm{L}$}
\newcommand{\ltot}            {L$_\mathrm{TOT}$}
\newcommand{\lmsx}            {L$_\mathrm{MSX}$}
\newcommand{\lsun}            {L$_\sun$}
\newcommand{\msun}            {M$_\sun$}

\newcommand{\tmb}             {$\mathrm{T_{mb}}$}
\newcommand{\avm}             {$\mathrm{M_v}$}
\newcommand{\novbv}           {$\mathrm{N_{O/B}}$}
\newcommand{\teff}            {$\mathrm{T_{eff}}$}

\shorttitle{NEAR-IR OBSERVATIONS OF UC~H~{\sc ii}\ REGIONS}
\shortauthors{ALVAREZ ET AL.}

\begin{document}


\title{NEAR-IR SUB-ARCSECOND OBSERVATIONS OF\\
    ULTRA-COMPACT H~{\sc ii} REGIONS\,\footnote{Based on observations at
    the ESO 3.6m telescope on La Silla observatory under Program-ID 
    No. 64.I-0606(B) and the 3.5m telescope on Calar Alto observatory 
    during the ALFA science verification program.}}


\author{\sc Carlos Alvarez, Markus Feldt, Thomas Henning, Elena Puga
    and Wolfgang Brandner} 
\affil{Max-Planck-Institut f\"ur Astronomie,
  K\"onigstuhl 17, D-69117 Heidelberg, Germany}
\and 
\author{\sc Bringfried Stecklum\altaffilmark{}}
\affil{Th\"{u}ringer Landessternwarte Tautenburg, 
  Sternwarte 5, D-07778 Tautenburg, Germany}








\begin{abstract}
We present adaptive-optics (AO) assisted $J$, $H$\ and $K/K'$\ images of 8
ultra-compact  H~{\sc ii}\ regions (\uchii{}s) taken with the ALFA 
and ADONIS AO systems at Calar Alto and La Silla observatories. The images show
details of the stellar population and the near-IR morphology of \uchii{}s
with unprecedented resolution. We have searched for the ionizing sources
of the regions using near-IR photometry. The spectral
type of the ionizing and most luminous stars inferred from our
photometry has been compared with spectral type estimates from IRAS
and published radio-continuum measurements. We find that the
photometric near-IR spectral types are earlier than estimates from radio
and IRAS data. This difference is alleviated when stellar spherical
models including line blanketing and stellar winds instead of non-LTE 
plane-parallel models are used to derive
the radio- and IRAS-based spectral types. We also made an attempt to 
correlate the properties of the near-IR ionizing population with MSX 
data and published CS measurements. No correlation was found. We note 
that in two of the regions (G309.92+0.48 and G61.48+0.09B1), the best 
candidate to ionize the region is possibly a super-giant. 
\end{abstract}



\keywords{H~\textsc{II} regions: general --- infrared: 
  ISM --- instrumentation: adaptive
  optics, high angular resolution --- catalogs}


\section[INTRODUCTION]{INTRODUCTION}
\label{Introduction}

The formation of massive stars ($M \ge 8 \, \mathrm{M}_{\odot}$) is a
major open problem in astrophysics. In contrast to the
formation of low-mass stars, the Kelvin-Helmholtz time scale for the
onset of nuclear fusion within massive protostars is shorter than the
accretion time scale. In other words, massive stars reach the Zero-Age
Main Sequence (ZAMS) while still being embedded in their natal
molecular clouds. It is commonly argued that during this phase the
radiation pressure is so high that it can substantially decrease or
even halt the infall (e.g \citealt{wolfire:1987}). However,
this scenario contradicts the existence of stars with
masses up to $100\, \mathrm{M}_{\odot}$.

Quite a number of loopholes from this dilemma have been proposed: mass
build-up through disc accretion (e.g. \citealt{mckee:2002,yorke:2002}), 
accretion with extremely large (e.g.
\citealt{nakano:2000}) or increasing (e.g. \citealt{behrend:2001})
infall rates, or -- as a completely different approach -- coalescence
of less massive stars in young and dense cluster environments
\citep{bonnell:98}.  Each model provides specific predictions which,
in principle, can be tested observationally, e.g.  the presence of
accretion discs, the existence of relations in the infall-outflow
dynamics, or an enhanced binarity/multiplicity frequency.

Observationally, however, we still know little about the early stages
of massive star formation. Massive stars are rare and hence
statistically located at larger distances than sites of low-mass star
formation. Additionally, massive stars tend to form in clusters
and associations. They have a strong impact on their environment
producing outflows, large and bright ionized regions, clumps of heated 
dust, and
reflection nebulae. To disentangle these complex, far-away regions,
very high sensitivity and resolution are required. Moreover, the natal
molecular clouds these objects are still embedded in, produce tens to
hundreds of magnitudes of visual extinction, which makes them
accessible only at long wavelengths.

To overcome these observational limitations, early surveys of sites of
massive star formation focused on using the Very Large Array (VLA) 
to detect the radio
continuum emitted from so-called ultra-compact H~{\sc ii} regions
(\uchii{}s). \uchii{}s represent a relatively evolved phase in the
young massive star's life, when it has already ionized a substantial
amount of surrounding gas (e.g. \citealt{wc89}). For long years, this
was the earliest phase of massive stars' lifes accessible to
astronomers, and only at wavelengths longer than 1\,mm.

Only since a few years, adaptive optics (AO) systems working in the
near-infrared (NIR) provide high enough resolution to start disentangling
individual stars. At these wavelengths, the visual extinction towards 
some of the embedded central ionizing stars can \emph{just} be
overcome. AO-assisted imaging of this kind allowed the identification 
of the ionizing objects and the stellar content in a number of sites 
of massive star formation \citep{feldt:98,feldt:99,henning:2001,Henning02}.

In this paper, we present the results of a mini survey of AO-assisted
NIR observations towards 8 \uchii{}s. The sample (see 
Table~\ref{TargetListTable}) was selected from the catalogues 
of \citet{Bronfman96}, \citet{wc89} and \citet{Kurtz94}. All sources
are located
within 30$''$\ of a bright optical star to be used as wavefront sensor
reference for the AO observations. 
We present the
photometry of sources found to be embedded in, or close to the known
sites of massive star formation. Basic stellar properties derived from 
the resulting colour information are also discussed. The stellar
population inferred from the near-IR photometry, and in particular, the
population of possible ionizing stars is compared with existing 
predictions based on radio, millimetre and mid-IR data.

\section[OBSERVATIONS AND DATA REDUCTION]{OBSERVATIONS AND DATA REDUCTION}
\label{Observations}

\subsection[ALFA Observations]{ALFA Observations}
\label{ALFA}

The observations using the AO system ALFA \citep{hippler:98} on the
3.5\,m telescope at the German-Spanish astronomical centre on Calar
Alto (Spain), were part of the ALFA science verification programme.
This programme was carried out from fall 1999 to fall 2000. The
data for the catalogue presented here were taken in the month of
September in both years. Individual dates are given in
Table~\ref{tab:ObsTable}. The science verification programme ensured
that the $K$-band seeing was always better than 1\arcsec, partly
reaching values as good as 0\farcs3. 

Omega-Cass \citep{lenzen:98} served as infrared camera.  The camera pixel
scale was 0\farcs04 for \goo\ and \gsvsv, and 0\farcs08 for the rest of
the targets. These scales result in a field of view (FOV) of 
40\arcsec$\times$40\arcsec and 80\arcsec$\times$80\arcsec, respectively.

The layout of the observations was generally as follows. The AO system
was locked onto the AO guide star and a first frame was taken. Then, the
telescope was moved to an offset position of about 5\arcsec - 10\arcsec\
with respect to the original position. The AO guide star was re-centred
on the wavefront sensor using the field-selecting mirror (FSM) of the
AO system. The AO loop was re-locked and another image taken with the 
same total integration time. This process was repeated for a total of 
5 dither positions to provide for a ``moving sky''. An overall
integration time of 10 minutes was achieved in each of the 3 filters 
($J$, $H$\ and $K'$).

\subsection[ADONIS Observations]{ADONIS Observations}
\label{ADONIS}

The ADONIS \citep{Beuzit94} observations were carried out in March
2000. The general
strategy was the same as for the ALFA observations. However, for the
dither pattern, it was not necessary to move the telescope, since for 
the ADONIS system, the FSM moves the field of view of the IR camera, 
instead of the wavefront sensor. The SHARP camera \citep{hoffmann:95} 
was used at a pixel scale of 0\farcs1. The images were taken in the $J$, 
$H$\ and $K$\ bands, with a total integration time of 10 minutes per 
filter.

\clearpage
   \begin{deluxetable}{lcrrrrrcl}
	\tablewidth{0pt}
	\tabletypesize{\scriptsize}
	\tablecaption{Target List \label{TargetListTable}}
	\tablehead{
	\colhead{Object} & \colhead{IRAS} & \colhead{$\alpha$ (\fh\,\fm\,\fs) \tablenotemark{a}} &
	\colhead{$\delta$ (\fdg\,\farcm\,\farcs) \tablenotemark{a}} &
	\colhead{$l$ (\fdg) \tablenotemark{b}} & \colhead{$b$ (\fdg) \tablenotemark{b}} & 
	\colhead{Ref \tablenotemark{c}} & \colhead{Type \tablenotemark{d}} & \colhead{Other Names}
	}
	\startdata
            \gton   & 13471-6120 & 13 50 41.8 & -61 35 11 & 309.92 & 0.48 & 1,2,3,7,9 & UN & GL\,4182 \\  
            \gtfo\tablenotemark{*}  & 17165-3554 & 17 19 58.2 & -35 57 32 & 351.16 & 0.70 & 1,4,7,9 & \nodata & NGC\,6334V \\  
            \gf     & 17574-2403 & 18 00 30.4 & -24 04 00 & 5.89 & -0.40 & 3,4,6,7,9,12 & SH & W28\,A2 \\  
            \goo    & 18085-1931 & 18 11 33.2 & -19 30 39 & 11.11 & -0.40 & 3,10,13 & IR & \nodata \\  
            \goe    & 18222-1317 & 18 25 01.0 & -13 15 40 & 18.15 & -0.29 & 3,4,10,13 & CO & \nodata \\  
            \gso\tablenotemark{**} & 19446+2505 & 19 46 46.6 & +25 12 31 & 61.47 & 0.09 & 3,4,7,10,12 & \nodata & Sh2-88B \\  
            \gsvo   & 19598+3324 & 20 01 45.6 & +33 32 44 & 70.29 & 1.60 & 3,4,5,6,7,8,13 & CH & K\,3-50\,A \\  
            \gsvsv  & 20277+3851 & 20 29 36.7 & +39 01 22 & 77.96 & -0.01 & 3,4,13 & IR & \nodata \\  
	\enddata
	\tablenotetext{a}{Right Ascension and Declination of the peak
	  radio emission taken from the literature 
	  (equinox J2000).} 
	\tablenotetext{b}{Galactic coordinates.} 
	\tablenotetext{c}{References for the position of the field of view.} 
	\tablenotetext{d}{Radio morphological type from Wood \&
            Churchwell (1989) Walsh et al. (1997), and Kurtz et
            al. (1994), depending on the source. UN - Unresolved/spherical, CO -
	    Cometary, CH - Core-halo, SH - Shell, IR - Irregular/multi-peaked.} 
	\tablenotetext{*}{Several compact radio continuum sources are included
	  in the field of view of our near-IR images.}
	\tablenotetext{**}{This source has two components. Both
	  components are included in the field of our near-IR images.
	  Component A classified as spherical, and component B as irregular.}
	\tablerefs{
          (1) Walsh et al. 1997;
	  (2) Braz et al. 1983;
          (3) Bronfman et al. 1996;
          (4) Lockman 1989;
          (5) Roelfsema et al. 1988;
          (6) Afflerbach et al. 1996;
          (7) Braz and Epchtein 1983;
	  (8) Blitz et al. 1982;
          (9) Walsh et al. 1998;
	  (10) Solomon et al. 1987;
          (11) Churchwell et al. 1978;
	  (12) Wood \& Churchwell 1989;
	  (13) Kurtz et al. 1994
	}
   \end{deluxetable}

\clearpage
       \begin{deluxetable}{lllllll} \tablewidth{0pt}
         \tabletypesize{\scriptsize}
	 \tablecaption{Observations and Data Quality\label{tab:ObsTable}}
         \tablehead{ \colhead{Object} & 
           \colhead{Date } &
           \colhead{Instrument / } &
           \colhead{Photometric} &
           \colhead{Bands} & 
           \colhead{SR / FWHM\tablenotemark{b}} & 
           \colhead{SR / FWHM}\\
           & 
           \colhead{Observed} &
           \colhead{Telescope\tablenotemark{a}} &
           \colhead{Calibrator} &
           \colhead{Available} &
           \colhead{(Guide Star)} &
           \colhead{(Target)}
           }
         \startdata
       \gton  & 2000 Mar 14 & ADONIS/ 3.6\,LS & 1719581-355732 & $J$, $H$, $K$ & -- / --\tablenotemark{c} & 0.03 / 0\farcs24\\
       \gtfo   & 2000 Mar 15  & ADONIS/ 3.6\,LS & 1719581-355732 & $J$, $H$, $K$ & -- / --\tablenotemark{c} & 0.14 / 0\farcs13\\
       \gf     & 2000 Sep 8 & ALFA / 3.5\,CA & 1800310-240409 & $J$, $H$, $K'$ & 0.04 / 0\farcs44 & 0.03 / 0\farcs55\\
       \goo    & 1999 Sep 22 & ALFA / 3.5\,CA & AS\,31-0 & $J$, $H$, $K'$ & 0.21 / 0\farcs16 & 0.02 / 0\farcs39 \\
       \goe    & 2000 Sep 9 & ALFA / 3.5\,CA & FS\,117 & $J$, $H$, $K'$ & 0.05 / 0\farcs48 & 0.01 / 0\farcs53 \\
       \gso    & 2000 Sep 16 & ALFA / 3.5\,CA & AS\,34-0 &  $J$, $H$, $K'$ & 0.10 / 0\farcs5 & 0.03 / 0\farcs57\\
       \gsvo   & 2000 Sep 10 & ALFA / 3.5\,CA & AS\,35 &  $J$, $H$, $K'$ & 0.20 / 0\farcs22 & 0.14 / 0\farcs22 \\
       \gsvsv  & 1999 Sep 22 & ALFA / 3.5\,CA & AS\,31-0 & $J$, $H$, $K'$ & 0.06 / 0\farcs31 & 0.02 / 0\farcs37\\
       \enddata 
       \tablenotetext{a}{3.6\,LS refers to the ESO 3.6\,m
	 telescope at La Silla, Chile.  3.5\,CA refers to
         the 3.5\,m telescope at Calar Alto, Spain}
       \tablenotetext{b}{Strehl ratio (SR) and full-width-half-maximum
	 (FWHM) of the PSF given at the corresponding $K$, or $K'$-band.}
       \tablenotetext{c}{The AO guide star is not within the FOV of
	 the IR image.}
   \end{deluxetable}

\clearpage

\subsection[Data Reduction]{Data Reduction}
\label{sec:DataReduction}

The data were sky-subtracted, flat-fielded and bad pixel-corrected
following the standard near-IR reduction procedures. Dome flat-fields 
were taken with three different levels of illumination, using always 
the same integration time. In this way, the response of each
individual pixel can be fitted against the median response of the 
detector, resulting in a good representation of small- and large-scale 
variations of pixel responses. Bad pixels were identified when the 
individual response was differing from 1.0 by more than a factor of
1.5.

A sky image was constructed for each object by median combination of
several dithered frames after being corrected from flatfield. Each
object frame was sky-subtracted, shifted and averaged to produce a 
final mosaic. For locations were a pixel was flagged as ``bad'' in one 
(or more) of the overlapping frames, only the values from the 
corresponding ``good'' frames were taken into account. The remaining
bad pixels -where no overlapping good frame could be found at all-
were corrected by interpolating between neighbouring pixels. 

\subsection[Data Quality]{Data Quality}
\label{sec:DataQuality}

Table \ref{tab:ObsTable} shows the quality of the data in terms of
resolution and Strehl ratio. Both numbers are given for the actual
targets and for the AO guide star (where available). The comparison
illustrates the general property of (classical) AO observation of
suffering from anisoplanatism, i.e. the image quality varies across the
field and it is generally worse on the targets than on the guide
stars. Our best-quality target is \gtfo, with a resolution of
0\farcs13 and a Strehl ratio of 0.14 (on-source). 

The FWHM was measured on unresolved sources only. When the target region
does not contain such a source, the nearest available unresolved star was used.
The number given corresponds to the average diameter of the PSF at
half the maximum intensity measured along profiles extracted at ten
different angles.

Determining the Strehl number is slightly more complicated, since it
requires the knowledge of the total flux of the unresolved source. 
Once the total flux is determined, a model PSF is created
from the telescope properties (diameters of primary and secondary
mirror, width of the spiders). The Strehl ratio is the ratio between 
the peak intensities of the measured and the modelled PSFs. For all
images taken with the ALFA system at a sampling of 0\farcs08 per 
pixel, the Strehl estimates should be seen as a lower limit due to 
the possibility of placing the PSF peak in between two pixels, which 
slightly reduces the peak flux. Generally, the Strehl numbers should be 
seen as estimates only due to the difficulty of determining total 
stellar flux levels in crowded regions.

\subsection[Photometry]{Photometry}
\label{sec:Photometry}

Rough photometric zero points were obtained by observing standard stars
immediately before and after each \uchii\ region. For some
targets, no separate standard stars were observed. In these cases, 
a point source inside the images was identified from the
2MASS\footnote{The Two Micron All Sky Survey is a joint project of 
the University of Massachusetts and the Infrared Processing and 
Analysis Center/California Institute of Technology, funded 
by the National Aeronautics and Space Administration and the National 
Science Foundation.} 
All-Sky Catalog of Point Sources \citep{Cutri03} and used for the flux
calibration. 
The names are also listed in column (4) of Table~\ref{tab:ObsTable}. 

Magnitudes were calculated using PSF-fitting photometry (DAOPHOT in
IRAF\footnote{IRAF is distributed by the National Optical Astronomy 
Observatories, which are operated by the Association of Universities 
for Research in Astronomy, Inc., under cooperative agreement with the 
National Science Foundation.}). The fitting radius was set to the
FWHM of the PSF. The PSF was best fitted with a Gaussian core and
Lorentzian wings. A quadratic variable PSF across the field of view
was used to account for the anisoplanatism of the AO images. When the
number of PSF stars available in the field of view was less than 9, 
(e.g. \gton) a linear variation of the PSF was used. In 
addition, aperture photometry was performed on isolated bright sources 
covering each field. The aperture was chosen to
be large enough to include the full PSF flux. The mean difference
between the magnitudes obtained with the aperture photometry and the
magnitudes resulting from PSF fitting photometry in the isolated
stars was utilised to calculate an aperture correction for each
image. The aperture correction was subtracted from the PSF-fitting
magnitudes, to determine the final software magnitudes. Errors in these 
magnitudes were calculated from the maximum between the error given 
by the PSF-fitting algorithm and the standard deviation of the aperture 
correction. In a further step, software magnitudes were transformed 
into 2MASS magnitudes to uniform the data set. Stars in common 
in our images and in the 2MASS All-Sky Catalog of Point Sources were 
used to determine the photometric transformation. Table~\ref{PhotometryTable} 
shows the final $J$, $H$\ and $K_s$\ magnitudes in the 2MASS system for 
some selected stars in each \uchii\ region. The final errors, shown 
in Table~\ref{PhotometryTable}, are the result of propagating the
errors in the software magnitudes and the errors from the linear fit 
used in the colour transformation. Limiting magnitudes in the range between 
18.5 and 19.5~mag at $J$, 15.8 and 17.9~mag at $H$, and 15.4 and 
16.8~mag at $K_s$\ were achieved.

\clearpage
\begin{deluxetable}{lrrrrrr}
  \tablewidth{0pt}
  \tablecaption{Photometry \label{PhotometryTable}}
  \tablehead{
    \colhead{Source} & \colhead{ID}  & \colhead{$\alpha$\,\tablenotemark{a}}  & \colhead{$\delta$\,\tablenotemark{a}}  & 
    \colhead{m$_{J}$\,\tablenotemark{b}} & \colhead{m$_{H}$\,\tablenotemark{b}} & 
    \colhead{m$_{K_s}$\,\tablenotemark{b}}\\
    \colhead{} & \colhead{} &
    \colhead{(h m s)}  & \colhead{($^\circ$\ $'$\ $''$)} & \colhead{(mag)}  & \colhead{(mag)} & 
    \colhead{(mag)} 
  } 
  \startdata
     \gton  &   3 & 13 50 42.97 & -61 34 56.2 & $ 9.6\pm0.1$ & $ 9.1\pm0.1$ & $ 9.0\pm0.2$ \\
            &   4 & 13 50 42.76 & -61 34 58.4 & $12.9\pm0.1$ & $12.3\pm0.1$ & $12.0\pm0.2$ \\
            &  12 & 13 50 42.36 & -61 35 08.0 & $14.5\pm0.1$ & $12.2\pm0.1$ & $11.2\pm0.2$ \\
 	    &  13 & 13 50 42.07 & -61 35 09.9 & $15.8\pm0.1$ & $13.7\pm0.1$ & $12.8\pm0.2$ \\
            &  14 & 13 50 41.84 & -61 35 10.6 &  \nodata     & $15.1\pm0.2$ & $11.3\pm0.2$ \\
 	    &  17 & 13 50 40.90 & -61 35 06.8 & $17.7\pm0.1$ & $13.8\pm0.1$ & $11.3\pm0.2$ \\
 	    &  39 & 13 50 41.78 & -61 35 11.5 & $18.7\pm0.2$ & $13.6\pm0.2$ & $10.2\pm0.2$ \\ \hline
     \gtfo  &  20 & 17 19 58.16 & -35 57 32.1 & $11.39\pm0.05$ & $11.2\pm0.2$ & $11.1\pm0.3$ \\ 
 	    &  24 & 17 19 58.10 & -35 57 44.8 & $15.45\pm0.05$ & $14.7\pm0.2$ & $14.2\pm0.3$ \\ 
            &  25 & 17 19 58.50 & -35 57 48.6 & $16.04\pm0.04$ & $15.5\pm0.2$ & $14.5\pm0.3$ \\ 
 	    &  27 & 17 19 57.67 & -35 57 41.7 & $13.99\pm0.06$ & $12.5\pm0.2$ & $11.8\pm0.3$ \\ 
 	    &  29 & 17 19 57.85 & -35 57 50.3 & $14.64\pm0.08$ & $12.7\pm0.2$ & $10.7\pm0.4$ \\ 
 	    &  33 & 17 19 57.22 & -35 57 26.4 &    \nodata   & $15.8\pm0.4$ & $13.4\pm0.5$ \\ 
 	    &  34 & 17 19 57.31 & -35 57 23.0 & $14.23\pm0.05$ & $13.3\pm0.2$ & $12.9\pm0.3$ \\
 	    &  36 & 17 19 57.19 & -35 57 20.7 &    \nodata   & $15.3\pm0.4$ & $13.4\pm0.3$ \\ 
 	    &  45 & 17 19 57.07 & -35 57 23.8 & $18.0\pm0.2$ & $12.7\pm0.4$ & $10.0\pm0.5$ \\
 	    &  46 & 17 19 56.85 & -35 57 27.3 & $15.9\pm0.1$ & $13.1\pm0.3$ & $11.4\pm0.4$ \\ \hline
     \gf    &   1 & 18 00 31.01 & -24 04 08.9 & $11.95\pm0.06$ & $10.8\pm0.2$ & $10.4\pm0.1$ \\
            &   3 & 18 00 30.90 & -24 04 02.5 & $17.27\pm0.07$ & $15.5\pm0.2$ & $14.3\pm0.1$ \\
 	    &   4 & 18 00 30.90 & -24 03 58.8 &    \nodata   & $17.2\pm0.2$ & $15.4\pm0.1$ \\
 	    &   5 & 18 00 30.87 & -24 04 04.1 & $16.39\pm0.08$ & $14.0\pm0.2$ & $13.0\pm0.1$ \\
 	    &   8 & 18 00 30.81 & -24 04 00.8 &    \nodata   & $16.9\pm0.2$ & $15.0\pm0.1$ \\
 	    &  12 & 18 00 30.59 & -24 04 02.0 & $15.99\pm0.06$ & $15.2\pm0.2$ & $14.8\pm0.1$ \\
 	    &  14 & 18 00 30.44 & -24 04 00.4 &    \nodata   & $16.2\pm0.2$ & $13.3\pm0.2$ \\
 	    &  16 & 18 00 30.69 & -24 03 56.0 &    \nodata   & $17.1\pm0.3$ & $12.1\pm0.3$ \\
 	    &  17 & 18 00 30.46 & -24 03 57.5 &    \nodata   & $16.0\pm0.2$ & $13.6\pm0.2$ \\
 	    &  20 & 18 00 30.31 & -24 04 00.2 & $18.3\pm0.1$ & $15.2\pm0.2$ & $13.0\pm0.2$ \\
 	    &  24 & 18 00 31.28 & -24 03 59.9 & $18.4\pm0.1$ & $17.1\pm0.2$ & $15.3\pm0.1$ \\
 	    &  25 & 18 00 31.36 & -24 04 08.7 & $18.3\pm0.1$ & $16.1\pm0.2$ & $14.8\pm0.1$ \\
 	    &  26 & 18 00 31.50 & -24 04 09.9 & $18.3\pm0.1$ & $16.6\pm0.2$ & $15.6\pm0.1$ \\
 	    &  70 & 18 00 29.91 & -24 04 06.3 & $17.07\pm0.08$ & $14.3\pm0.2$ & $12.9\pm0.1$ \\
 	    &  76 & 18 00 29.31 & -24 03 57.1 & $10.85\pm0.06$ & $10.5\pm0.2$ & $10.2\pm0.1$ \\
            &  81 & 18 00 32.35 & -24 04 27.8 & $16.9\pm0.1$ & $12.2\pm0.2$ & $ 9.8\pm0.2$ \\
 	    &  82 & 18 00 32.32 & -24 04 26.2 & $16.08\pm0.08$ & $13.1\pm0.2$ & $11.8\pm0.1$ \\
 	    &  83 & 18 00 32.14 & -24 04 27.3 & $18.6\pm0.2$ & $13.5\pm0.2$ & $11.7\pm0.1$ \\
            &  86 & 18 00 32.33 & -24 04 09.5 & $18.4\pm0.1$ & $16.2\pm0.2$ & $13.9\pm0.2$ \\
 	    & 120 & 18 00 31.65 & -24 04 14.9 & $14.40\pm0.07$ & $12.1\pm0.2$ & $11.0\pm0.1$ \\ \hline
     \goo   &   2 & 18 11 32.35 & -19 30 52.8 & $13.3\pm0.2$ & $10.7\pm0.1$ & $ 9.6\pm0.2$ \\  
 	    &   3 & 18 11 32.20 & -19 30 49.7 & $15.2\pm0.1$ & $14.1\pm0.1$ & $13.8\pm0.1$ \\  
 	    &   4 & 18 11 31.54 & -19 30 41.4 & $17.4\pm0.2$ & $14.9\pm0.1$ & $14.1\pm0.2$ \\  
 	    &   8 & 18 11 32.07 & -19 30 42.7 &    \nodata   & $17.3\pm0.2$ & $14.6\pm0.4$ \\  
 	    &   9 & 18 11 31.98 & -19 30 40.9 &    \nodata   & $16.3\pm0.2$ & $14.6\pm0.3$ \\  
 	    &  10 & 18 11 32.71 & -19 30 44.8 & $16.4\pm0.2$ & $14.2\pm0.1$ & $13.1\pm0.2$ \\  
 	    &  11 & 18 11 32.65 & -19 30 43.7 & $14.1\pm0.1$ & $13.4\pm0.1$ & $13.2\pm0.1$ \\  
 	    &  12 & 18 11 31.85 & -19 30 38.7 & $18.6\pm0.4$ & $14.1\pm0.2$ & $12.3\pm0.3$ \\  
 	    &  15 & 18 11 32.18 & -19 30 38.3 & $16.5\pm0.1$ & $15.5\pm0.1$ & $15.5\pm0.1$ \\  
 	    &  22 & 18 11 31.40 & -19 30 28.5 &    \nodata   & $15.8\pm0.2$ & $13.2\pm0.5$ \\  
 	    &  33 & 18 11 33.14 & -19 30 31.3 & $ 9.4\pm0.1$ & $ 9.05\pm0.09$ & $ 9.0\pm0.1$ \\ \hline
     \goe   &  31 & 18 25 01.96 & -13 16 13.4 & $14.22\pm0.08$ & $13.0\pm0.1$ & $12.44\pm0.09$ \\	
 	    &  49 & 18 25 01.82 & -13 16 10.5 & $16.08\pm0.09$ & $14.3\pm0.1$ & $13.2\pm0.1$ \\
 	    &  59 & 18 25 00.93 & -13 16 06.8 & $17.0\pm0.1$ & $14.9\pm0.1$ & $13.7\pm0.1$ \\
 	    &  63 & 18 25 01.21 & -13 16 03.1 & $14.52\pm0.08$ & $13.2\pm0.1$ & $12.71\pm0.09$ \\
 	    &  79 & 18 25 00.14 & -13 15 56.1 & $16.2\pm0.1$ & $14.0\pm0.2$ & $13.2\pm0.1$ \\
 	    &  85 & 18 25 00.40 & -13 15 54.9 & $11.90\pm0.07$ & $11.1\pm0.1$ & $10.80\pm0.09$ \\
 	    &  92 & 18 25 02.24 & -13 15 54.3 & $17.4\pm0.1$ & $15.2\pm0.2$ & $14.3\pm0.1$ \\
 	    & 113 & 18 25 01.15 & -13 15 48.9 & $ 9.35\pm0.06$ & $ 8.8\pm0.1$ & $ 8.71\pm0.09$ \\
 	    & 119 & 18 25 01.71 & -13 15 46.4 & $15.06\pm0.09$ & $13.5\pm0.1$ & $12.8\pm0.1$ \\
 	    & 120 & 18 25 00.58 & -13 15 45.5 & $17.5\pm0.1$ & $15.4\pm0.2$ & $14.2\pm0.1$ \\
 	    & 121 & 18 25 01.12 & -13 15 44.4 & $15.0\pm0.1$ & $13.1\pm0.1$ & $11.9\pm0.1$ \\
 	    & 143 & 18 25 01.65 & -13 15 41.4 & $12.23\pm0.08$ & $10.8\pm0.1$ & $10.09\pm0.09$ \\
 	    & 144 & 18 25 01.12 & -13 15 40.1 & $16.8\pm0.1$ & $13.5\pm0.2$ & $11.6\pm0.1$ \\
 	    & 147 & 18 25 00.50 & -13 15 37.9 & $15.47\pm0.07$ & $14.5\pm0.1$ & $13.85\pm0.09$ \\
 	    & 153 & 18 25 01.99 & -13 15 32.9 & $12.69\pm0.06$ & $12.1\pm0.1$ & $11.84\pm0.09$ \\
 	    & 155 & 18 24 58.38 & -13 15 29.7 & $14.6\pm0.1$ & $12.4\pm0.2$ & $11.7\pm0.1$ \\
 	    & 159 & 18 24 59.09 & -13 15 27.5 & $17.7\pm0.1$ & $14.8\pm0.2$ & $13.5\pm0.1$ \\
            & 169 & 18 25 02.33 & -13 15 19.6 & $12.48\pm0.08$ & $11.3\pm0.1$ & $10.95\pm0.09$ \\
 	    & 173 & 18 25 01.43 & -13 15 17.9 & $15.2\pm0.1$ & $13.3\pm0.1$ & $12.4\pm0.1$ \\
 	    & 175 & 18 25 00.17 & -13 15 14.7 & $12.98\pm0.08$ & $11.7\pm0.1$ & $11.38\pm0.09$ \\ \hline
     \gso   &  13 & 19 46 49.13 &  25 12 07.2 & $15.7\pm0.1$ & $13.54\pm0.08$ & $12.2\pm0.1$ \\
 	    &  19 & 19 46 46.92 &  25 12 13.4 & $10.3\pm0.1$ & $ 9.65\pm0.08$ & $ 9.34\pm0.09$ \\
 	    &  50 & 19 46 47.83 &  25 12 30.2 & $16.6\pm0.1$ & $14.31\pm0.09$ & $12.7\pm0.1$ \\
 	    &  61 & 19 46 47.12 &  25 12 34.1 & $14.7\pm0.1$ & $13.01\pm0.08$ & $12.1\pm0.1$ \\
 	    &  82 & 19 46 47.60 &  25 12 45.6 & $16.1\pm0.1$ & $12.2\pm0.1$ & $ 9.4\pm0.2$ \\
 	    &  83 & 19 46 47.32 &  25 12 45.6 & $13.4\pm0.1$ & $11.84\pm0.08$ & $10.8\pm0.1$ \\
 	    &  84 & 19 46 47.05 &  25 12 45.8 & $14.4\pm0.1$ & $13.17\pm0.08$ & $12.3\pm0.1$ \\
 	    & 111 & 19 46 47.29 &  25 12 59.8 & $15.9\pm0.1$ & $13.52\pm0.09$ & $12.0\pm0.1$ \\
 	    & 112 & 19 46 49.05 &  25 13 00.7 & $11.9\pm0.1$ & $11.35\pm0.08$ & $11.07\pm0.09$ \\
 	    & 116 & 19 46 48.35 &  25 13 02.9 & $18.2\pm0.1$ & $14.83\pm0.09$ & $12.7\pm0.1$ \\ \hline
     \gsvo  &  11 & 20 01 46.93 &  33 32 52.2 & $15.82\pm0.06$ & $14.82\pm0.06$ & $14.5\pm0.1$ \\
            &  29 & 20 01 46.32 &  33 32 35.1 & $13.64\pm0.06$ & $12.86\pm0.06$ & $12.7\pm0.1$ \\
 	    &  47 & 20 01 45.98 &  33 32 37.6 & $19.4\pm0.1$ & $17.11\pm0.07$ & $15.6\pm0.1$ \\
 	    &  52 & 20 01 45.61 &  33 32 32.7 & $19.09\pm0.09$ & $16.17\pm0.08$ & $14.4\pm0.1$ \\
 	    &  67 & 20 01 45.87 &  33 32 43.7 & $16.66\pm0.06$ & $15.22\pm0.07$ & $13.7\pm0.1$ \\
 	    &  68 & 20 01 45.69 &  33 32 43.4 & $15.96\pm0.09$ & $12.46\pm0.09$ & $ 9.1\pm0.2$ \\
 	    &  76 & 20 01 44.95 &  33 32 38.4 & $ 9.73\pm0.06$ & $ 9.14\pm0.06$ & $ 9.1\pm0.1$ \\
 	    & 126 & 20 01 42.30 &  33 32 37.7 & $12.86\pm0.06$ & $12.05\pm0.06$ & $11.9\pm0.1$ \\
 	    & 137 & 20 01 42.48 &  33 32 20.8 & $13.75\pm0.06$ & $13.33\pm0.06$ & $13.3\pm0.1$ \\
 	    & 181 & 20 01 45.75 &  33 32 24.7 & $18.25\pm0.08$ & $15.28\pm0.08$ & $13.8\pm0.1$ \\
 	    & 198 & 20 01 44.49 &  33 32 03.2 & $11.22\pm0.06$ & $10.58\pm0.06$ & $10.5\pm0.1$ \\ \hline
     \gsvsv &   4 & 20 29 37.29 &  39 01 18.5 & $17.7\pm0.1$ & $16.29\pm0.05$ & $14.41\pm0.09$ \\
 	    &   7 & 20 29 36.95 &  39 01 22.5 & $12.07\pm0.08$ & $10.84\pm0.05$ & $10.21\pm0.06$ \\
            &   9 & 20 29 36.78 &  39 01 22.6 & $15.5\pm0.2$ & $13.99\pm0.04$ & $12.94\pm0.05$ \\ 
            &  10 & 20 29 36.90 &  39 01 26.0 & \nodata        & $15.83\pm0.06$ & $14.33\pm0.08$ \\ 
            &  11 & 20 29 36.66 &  39 01 22.0 & \nodata        & $15.38\pm0.06$ & $12.27\pm0.05$ \\ 
 	    &  16 & 20 29 36.54 &  39 01 04.4 & $17.09\pm0.09$ & $14.28\pm0.07$ & $12.94\pm0.08$ \\
 	    &  19 & 20 29 35.54 &  39 00 54.5 & $14.66\pm0.09$ & $12.50\pm0.06$ & $10.84\pm0.09$ \\
 	    &  23 & 20 29 35.97 &  39 01 12.7 & $ 9.16\pm0.08$ & $ 8.46\pm0.05$ & $ 8.17\pm0.06$ \\
 	    &  30 & 20 29 35.09 &  39 01 10.5 & $10.88\pm0.08$ & $10.74\pm0.05$ & $10.59\pm0.06$ \\
            &  45 & 20 29 37.38 &  39 01 13.8 & $12.08\pm0.08$ & $11.01\pm0.05$ & $10.38\pm0.06$ \\ 
  \enddata
  \tablenotetext{a}{Right ascension and declination in equinox J2000.}
  \tablenotetext{b}{Magnitudes in the 2MASS photometric system.}
\end{deluxetable}

\clearpage

\subsection[Astrometry]{Astrometry}
\label{sec:Astrometry}

Astrometry was performed by matching the positions of stars in
common in our images and in the 2MASS survey images. The astrometric
accuracy in our images is the result of propagating the error from the
fit to obtain the plate solution, and the absolute astrometric
accuracy of the 2MASS catalogue. The first term ranges from
0\farcs03 to 0\farcs3 depending on the number of stars that
were available to perform the fit (never less than 4). The second term is 
0\farcs07 - 0\farcs08 following the explanatory supplement of the
2MASS survey. Hence our astrometric accuracy ranges between 0\farcs08
and 0\farcs3. Positions of selected stars are given in columns~(3) 
and~(4) of Table~\ref{PhotometryTable}.




\section[PHYSICAL PARAMETERS]{PHYSICAL PARAMETERS}
\label{PhysicalParameters}

\subsection[Distances]{Distances}
\label{Physics:Distances}

Table~\ref{DistancesTable} shows the kinematical distances to our
sources. The radial velocities were obtained from the literature. The
velocity tracers used were CO, CS, \methanol\ and \water\ masers, and
H~{\sc i}\ radio recombination lines. The Galactic rotation curve given by 
\citet{WouterlootBranz89}
was applied ($\mathrm{\theta(R)=\theta_\sun(R_\sun)^{0.0362}}$). The values 
R$_\sun$\,=\,8.5~kpc and $\theta_\sun$\,=\,220\kms\ were used for the 
distance from the Sun to the Galactic Centre and the tangential solar
velocity, respectively. 

The calculated values for the distance to the Sun
(\dsun) and the distance to the galactic centre (\dgal) are listed
in Table~\ref{DistancesTable}. In some cases, the distances quoted in
the literature are slightly different from the values shown here. This can be 
due, for instance, to the use of a different Galactic rotation curve. 
For the sources within the solar circle, we chose always the
solution of the near distance, since the solution of the far distance 
would lead to unrealistic (over-luminous) spectral types for most of the 
stars in the field of view.

For all sources but \goo\ and \gso, an average of the values given in
Table~\ref{DistancesTable} was adopted as the distance to the region, 
since all velocity tracers are in reasonable agreement with each
other. We used the standard deviation
of the average as the error in the distance. For \goo, our calculation 
of the kinematical distance yields only one value, 17~kpc. The
adoption of this distance would imply unrealistically bright
magnitudes for most of the stars in the field of view of \goo. Hence, 
we use the distance of 5.2~kpc from \citet{Kurtz94}. For \gso, we
found rather discrepant velocities depending on 
the tracers. We adopt a distance of 2.7~kpc, which is close to 
the most accepted values \citep[see ][]{Deharveng00}. 


\subsection[Ionizing Sources]{Ionizing Sources}
\label{Physics:IonizingSources}

\subsubsection[Near-IR Photometry]{Near-IR Photometry}
\label{Physics:IonizingSources:NIRPhotometry}

The definition of a near-IR source as a possible ionizing source of the
\uchii\ region is somewhat dependent on the region. This is in part 
due to the fact that some \hii\ regions, defined as {\em ultra-compact} in 
a high-resolution configuration of the VLA, are in fact extended   
over a few parsecs \citep[e.g.][]{Kurtz99,KimKoo01} at lower spatial 
resolution configurations, which are more sensitive to larger spatial
scales. Uncertainties in the distance and in the spectral type
derived from our photometry also affect the determination of sources 
possibly ionizing the \uchii\ region. We consider a source to be candidate
for ionizing an \hii\ region when it is located within a projected distance of 
0.5~pc from the radio-emission peak. This value corresponds to the
upper limit for the size of compact \hii\ regions. Another condition
is that the spectral type inferred from our photometry should be
earlier than B5V. However, these requirements 
are not sufficient. Potential ionizing sources should appear as point-like 
in our near-IR images. There should also be features in the images
that link them with the radio peak, e.g. near-IR
nebulosities. We discard as possible ionizing sources any
bright stars whose colours and low extinction indicate their being
foreground stars (see Sec.~\ref{MassfunctionSec}). Finally, we
associate a spectral type to each source in the FOV based on its
location in colour-colour (C-C) (\jh\ vs. \hks) and colour-magnitude 
(C-M) (\ks\ vs. \hks) diagrams. 

\clearpage
   \begin{deluxetable}{lrrcc@{\hspace{0.3cm}}cc}
	\tablewidth{0pt}
	\tablecaption{Distance Estimates \label{DistancesTable}}
	\tablehead{
	\colhead{Object} & \colhead{V$_\mathrm{LSR}$ (\kms)} & Tracer
	& \colhead{\dgal (kpc)\,\tablenotemark{a}} &
	\multicolumn{2}{c}{\dsun (kpc)\,\tablenotemark{a}}  & \colhead{Ref\,\tablenotemark{b}}
	}
	\startdata
            \gton\tablenotemark{c}  & -58.4 & CS           & 6.5  &  5.5  &   \nodata   & 3 \\
                                    & -60   & \methanol    & 6.5  &  5.5  &   \nodata   & 1 \\
                                    & -74   & \water       & 6.5  &  5.5  &   \nodata   & 7 \\
                                    & -59.9 & \methanol    & 6.5  &  5.5  &   \nodata   & 9 \\
            \gtfo                   & -5    & \methanol    & 7.3  &  1.2  & (15.6) & 1 \\  
                                    & -3.4  & \halphaef    & 7.7  &  0.8  & (16.0) & 4 \\  
                                    & -6    & \water       & 7.2  &  1.3  & (15.5) & 7 \\  
                                    & -6.3  & \methanol    & 7.1  &  1.4  & (15.4) & 9 \\  
            \gf                     & 9.3   & CS           & 5.9  &  2.6  & (14.3) & 3 \\  
                                    & 10.1  & \halphaooo   & 5.8  &  2.7  & (14.2) & 4 \\  
                                    & 5.0   & \halphasvs   & 6.9  &  1.6  & (15.3) & 6 \\  
                                    & 14    & \water       & 5.1  &  3.4  & (13.5) & 7 \\  
                                    & 9.6   & \water       & 5.9  &  2.6  & (14.3) & 9 \\  
            \goo                    & -1.1  & CS           & 8.8  &  \nodata   &  16.9  & 3 \\  
                                    & -2    & CO           & 9.0  &  \nodata   &  17.1  & 10 \\  
            \goe                    & 54.9  & CS           & 4.6  &  4.3  & (11.9) & 3 \\  
                                    & 53.9  & \halphaef    & 4.6  &  4.3  & (11.9) & 4 \\  
                                    & 54    & CO           & 4.6  &  4.3  & (11.9) & 10 \\  
                                    & 53.5  & \halphaoon   & 4.7  &  4.2  & (11.9) & 11 \\  
            \gso\tablenotemark{d}   & 21.9  & CS           & 7.6  &  2.7  &  (5.5) & 3 \\  
                                    & 27.3  & \halphaef    & 7.5  &  4.1  &   \nodata   & 4 \\  
                                    & 30    & \water       & 7.5  &  4.1  &   \nodata   & 7 \\  
                                    & 22    & CO           & 7.6  &  2.7  &  (5.5) & 10 \\  
            \gsvo                   & -25.2 & CS           & 9.7  &   8.4  &  \nodata  & 3 \\  
                                    & -24.5 & \halphaef    & 9.7  &   8.3  &  \nodata  & 4 \\  
                                    & -27.5 & \halphaonono & 9.8  &   8.4  &  \nodata  & 5 \\  
                                    & -19   & \water       & 9.4  &   7.8  &  \nodata  & 7 \\  
                                    & -24.5 & CO           & 9.7  &   8.3  &  \nodata  & 8 \\  
            \gsvsv                  & -2.9  & CS           & 8.6  &   4.1  &  \nodata  & 3 \\  
                                    & -5.5  & \halphaef    & 8.7  &   4.4  &  \nodata  & 4 \\  
	\enddata
	\tablenotetext{a}{Kinematical distances to the Galactic Centre
	  (column (4)) and to the Sun (column (5)) calculated from the
          radial velocities in column~(2). For the sources within the solar
          circle, the solution for the far distance is shown in brackets.}
	\tablenotetext{b}{References for the local-standard-rest
          velocity (V$_\mathrm{LSR}$) and velocity tracers.} 
	\tablenotetext{c}{The tanget point was used, since no solution 
	  was found.}
	\tablenotetext{d}{The tanget point was used for \halphaef\ and
	  \water, since not solution was found.}
	\tablerefs{
          (1) Walsh et al. 1997;
	  (2) Braz et al. 1983;
          (3) Bronfman et al. 1996;
          (4) Lockman 1989;
          (5) Roelfsema et al. 1988;
          (6) Afflerbach et al. 1996;
          (7) Braz \& Epchtein 1983;
	  (8) Blitz et al. 1982;
          (9) Walsh et al. 1998;
	  (10) Solomon et al. 1987;
          (11) Churchwell et al. 1978
	}
   \end{deluxetable}

\clearpage

In each diagram, we plot the theoretical Main Sequence (MS) and the 
giant and super-giant branches, at the assumed distance for each \uchii\
region. Intrinsic stellar colours were taken from  
\citet{Tokunaga00}\,\footnote{Available at \url{http://www.jach.hawaii.edu/JACpublic/UKIRT/}}.
The earliest dwarf for which intrinsic colours are available has a
spectral type O6V, and therefore, in the C-M diagrams we plot the 
Main Sequence up to this spectral type. For the giant and super-giant 
branches, we also plot only the spectral-type ranges available in
\citet{Tokunaga00}. \citet{Ducati01} show more recent stellar 
intrinsic colours, but they do not reach dwarf spectral types earlier 
than B0V. We therefore adopt the colours from Tokunaga rather than those from 
Ducati et al. for consistency, since the later lack of spectral 
types for OV stars. 

Absolute visual magnitudes were taken from \citet{Aller82}. The
association between \avm\ and the intrinsic colours was made by 
matching the spectral types of \citet{Aller82} with those of 
\citet{Tokunaga00}. Hence, we used the calibration of the 
\avm\,--\,spectral type relation given in \citet{Aller82}. Other 
possible calibrations of \avm\ as a function of the spectral type are
available in \citet{Vacca96} and \citet{Smith02}. For 
early type stars (O3,O4,O5), a calibration is also available in
\citet{Crowther98a}. However, the latter three were not used in 
our C-C and C-M diagrams, since they only have a limited
spectral-type coverage compared with the calibration by
\citet{Aller82}. The typical
errors associated to the \avm\ are $\sim$\,0.5~mag 
\citep[see][]{Vacca96}, which is larger than the typical error in 
our photometry. Hence, our spectral type classification 
based on near-IR photometry is not strongly affected by the  
\avm\,--\,spectral type relation. 

The following procedure was used to determine spectral
types based on the near-IR photometry. From the location 
of a star in the C-C diagram with respect to the unreddened Main
Sequence, its intrinsic colour excess and extinction were estimated. 
In most of the cases, sources with intrinsic near-IR excess appear 
to the right of the reddened MS. This \hks\ excess was assumed to 
be only due to a \ks\ excess{\bf, probably due to emission lines
  associated with photoionization and/or wind shocks, and/or continuum
  excess due to accretion luminosity. Hence, we consider this excess
  independent of the extinction.} The \ks\ excess was evaluated by
measuring the distance along the line \jh\ constant for the star 
between the point representing the star in the C-C diagram and 
the extincted Main Sequence. The extinction towards this star 
was then obtained by measuring the length of the reddening vector 
between the unreddened MS and the line \jh\ constant for the star. 
Once intrinsic \ks\ excess and reddening were known 
for a given star,  its \ks\ excess was firstly removed along the \hks\ and
\ks\ axes in the C-M diagram, to obtain a new point representing the 
star without intrinsic 
near-IR excess. Secondly, this point was projected onto the
unreddened MS (and giant and super-giant branches) following a 
line parallel to the reddening vector. The spectral type was then 
read directly from the unreddened MS (and giant and super-giant 
branches).

This method of spectral type determination, even though it is 
rather qualitative, it is reasonable for the given the typical 
uncertainties associated with the distances and with our measurements. 
A 25\% decrease in the distance, for instance, implies a shift 
of the MS of about 0.5~mag towards brighter magnitudes, which 
in turn implies a shift of $\sim$\,5 sub-types in the  
spectral type classification of a star. Typical errors in our
measurements are slightly smaller yielding similar uncertainties 
in the classification. Other sources of uncertainty are due to 
the \avm\,--\,spectral type relation used ($\sim$\,0.5 mag, 
see above) and due to the assumption of a universal extinction 
law. The latter is difficult to estimate, but different extinction 
laws towards each region due to different dust properties, would 
yield changes on the reddening vector slope in the C-C and C-M 
diagrams, which would affect the spectral type determination. 

\clearpage
\begin{deluxetable}{llrcccccccc}
  \tabletypesize{\scriptsize}
  \tablewidth{0pt}
  \tablecaption{Physical Parameters \label{PhysicalParametersTable}}
  \tablehead{
    \colhead{Object} & \colhead{D$_\sun$\,\tablenotemark{a}} &
    \colhead{Sp\,Ty\,\tablenotemark{b}} & \colhead{log(\nlyc)\,\tablenotemark{c}} & \colhead{Refs\,\tablenotemark{d}}  &
    \colhead{Sp\,Ty\,\tablenotemark{e}} & \colhead{Sp\,Ty\,\tablenotemark{f}} & \colhead{\ltot\,\tablenotemark{g}} & \colhead{Sp\,Ty\,\tablenotemark{e}} & \colhead{Sp\,Ty\,\tablenotemark{f}} & \colhead{\lmsx\,\tablenotemark{h}} \\
    \colhead{} & \colhead{(kpc)} &
    \colhead{(This work)}  & \colhead{} & \colhead{}  & \colhead{(Radio)} & \colhead{(Radio)} & \colhead{($10^4$\ L$_\sun$)} & \colhead{(IRAS)} & \colhead{(IRAS)} & \colhead{($10^4$\ L$_\sun$)} 
  } 
  \startdata
  \gton                  & 5.5\tablenotemark{**} & $>$O6V/OI & 48.0  & (1) & B0V      & O9V      &    32$\pm$2   & O6.5V      & O5V      &   11$\pm$3   \\
  \gtfo\tablenotemark{*} & 1.2$\pm$0.2           & O6V       & 47.3  & (2) & $<$B0.5V & B0V      &     4$\pm$1   & $<$B0.5V   & $<$B0V   &  0.9$\pm$0.4 \\
  \gf                    & 2.6$\pm$0.6           & O3V\tablenotemark{***}       & 48.6  & (3) & O9V      & O8V      &    27$\pm$13  & O7V        & O7V      &    5$\pm$3   \\
  \goo                   & 5.2\tablenotemark{**} & O6V       & 47.8  & (4) & $<$B0.5V & O9V      &   7.4$\pm$0.8 & B0V        & O9V      &  1.4$\pm$0.4 \\
  \goe                   & 4.3$\pm$0.1           & O6V       & 46.8  & (4) & $<$B0.5V & B0.5V    &    23$\pm$2   & O7V        & O7V      & 0.71$\pm$0.04\\
  \gso{}B1               & 2.7\tablenotemark{**} & O9I       & 48.4  & (5) & O9.5V    & O8V      &    17$\pm$1   & O8V        & O7.5V    &    5$\pm$1   \\
  \gso{}B2               & 2.7\tablenotemark{**} & B0V       & 46.5  & (3) & $<$B0.5V & B0V      &    17$\pm$1   & O8V        & O7.5V    &    5$\pm$1   \\
  \gsvo                  & 8.2$\pm$0.2           & $>$O6V/A0I& 49.2  & (4) & O6.5V    & O5V      &   200$\pm$16  & $>$O3V     & $>$O3V   &   52$\pm$15  \\
  \gsvsv                 & 4.2$\pm$0.2           & O8V       & 46.5  & (4) & $<$B0.5V & B1V      &   9.6$\pm$1   & O9.5V      & O8V      &    4$\pm$1   \\
  \enddata	
  \tablenotetext{a}{Average distance from
    Table~\ref{DistancesTable}. The errors 
    correspond to the standard deviation of the mean.} 
  \tablenotetext{b}{Spectral type of the best candidate for
    ionizing source from our near-IR data.The symbol ``$<$''
    denotes ``\emph{later than}'' spectral types. The symbol 
    ``$>$'' denotes ``\emph{earlier than}'' spectral types.} 
  \tablenotetext{c}{\nlyc\ is the number of Lyman continuum
    photons in units of s$^{-1}$ at the distance of column (2), 
    inferred from the radio continuum data shown in the 
    references of column (6).}	
  \tablenotetext{d}{References for the radio source used to
    determine spectral type.} 
  \tablenotetext{e}{Spectral type using the model grid of
    \citet{Vacca96}. A caveat should be noted here due to large
      IRAS beam size.} 
  \tablenotetext{f}{Spectral type using the model grid of
    \citet{Smith02}. A caveat should be noted here due to large 
      IRAS beam size.}
  \tablenotetext{g}{Total luminosity in units of $10^4$ \lsun\
    used to determine the spectral type. \ltot\ is calculated
    from the IRAS fluxes extracted from the IRAS PSC.}
  \tablenotetext{h}{MSX luminosity in units of $10^4$ \lsun\
    calculated from the integrated flux in a black-body
    fit to the MSX fluxes at 14.65~\micron\ and 21.34~\micron.}
  \tablenotetext{*}{Complex of compact radio-continuum
    sources. The radio spectral determination is from
    the large-scale shell structure \citep{Jackson99}.} 
  \tablenotetext{**}{No statistical error is given for \gton,
    since all velocity tracers yield the same distance. For 
    \goo\ and \gso, the most accepted distance found in the 
    literature is adopted, hence no error is given either.} 
  \tablenotetext{***}{Spectral type from \citet{Feldt03} scaled to our
    adopted distance.} 
  \tablerefs{
    (1) Walsh et al. 1998; (2) Jackson
    \& Kraemer 1999; (3) Wood \& Churchwell 1989; (4) Kurtz et
    al. 1994; (5) Deharveng et al. 2000
  }
\end{deluxetable}

\clearpage

\subsubsection[Radio]{Radio}
\label{Physics:InonisingSourcesRadio}

In Table~\ref{PhysicalParametersTable}, we give the 
spectral type of the ionizing star derived from radio-continuum 
observations found in the literature, taking our distances into
account. For half of regions but (\gf, \goe, \gso, and \gsvo),  
the spectral type given by the authors in the references of column~(5) 
was transformed into a Lyman continuum photon rate (\nlyc) using 
the same stellar model grids as those adopted by the original authors
(normally, \citealp{Panagia73} or \citealp{Vacca96}). This number 
was scaled to our assumed distance (column~(2)) for each source.  
The scaled log(\nlyc) is listed in column~(4). The log(\nlyc)\,--\,spectral 
type relation given in the stellar model grids from \citet{Vacca96} 
and \citet{Smith02} was used to infer our new 
estimate of the spectral type for the ionizing source (columns~(6) and
(7), respectively). In the case of \gton, \gtfo\ and \gsvsv, for which 
there are no previous estimates of the Lyman photon rate, the \nlyc\ was
calculated directly from the radio flux densities given in \citet{Walsh98}, 
\citet{Jackson99} and \citet{Kurtz94}, respectively. Eqs.~(1) and (3)
in \citet{Kurtz94} for optically thin \hii\ regions were applied. 
For \goo, we also used these equations to estimate the log(\nlyc) from
the radio flux density given in \citet{Kurtz94}. In these
four sources, the same log(\nlyc)\,--\,spectral type relations as for the 
rest of the \uchii{}s was applied to determine the spectral type of 
the ionizing source. No error is given in the log(\nlyc) for any of the
\uchii{}s because the original references for the radio data do not 
provide any error in the \nlyc\ or in the integrated radio
fluxes. An error based only on the statistical uncertainties 
would be negligible compared with the systematic errors 
implicit in the equations used to calculate the log(\nlyc) (e.g 
distance, assumption of optically thin emission or assumption 
of a geometry).

\subsection[IRAS and MSX Luminosities]{IRAS and MSX Luminosities}
\label{Physics:IRASMSX}

In Table~\ref{PhysicalParametersTable}, we also list the spectral
type deduced from the mid- and far-IR luminosity of the IRAS source
associated with each \uchii\ region. In all sources, we firstly apply the 
standard procedure to estimate the total luminosity (\ltot) based on 
the IRAS fluxes taken from the IRAS Point Source Catalogue 
(PSC, version 2.0). This consists of adding the total flux in each of the 
IRAS bands using 
the formula given in \citet{Walsh97} (see also \citealp{Henning90}). 
The total flux was divided by a correction factor that accounts for the 
flux at longer wavelengths than the IRAS 100-\micron\ band. We assumed 
a value of 0.61$\pm$0.02 for this factor \citep{Walsh97}. The 
luminosity (\ltot\ in column (8)) was calculated at the distances listed in 
column~(2). This luminosity was then used to derive a spectral 
type (columns (9) and~(10)), assuming one single heating star.  The 
spectral type for a given \ltot\ was read directly off Tables~5, 6
and~7 in \citet{Vacca96} to yield the entries of column (9). From
the Smith et al. grid, the \ltot\ was firstly transformed into a 
stellar effective temperature (\teff) by using the grid of 
\citet{Vacca96}. Secondly, the \teff\,--\,spectral type relation given 
in \citet{Smith02} was used to derive the spectral type.
One of the main drawbacks of this spectral type classification is that
the IRAS beam is $\sim$\,2$'$\ in size, i.e. it covers completely the 
field of view of our images. Hence, the assumption that only one star is 
contributing to the total luminosity derived from IRAS fluxes should be 
taken with extreme caution in the majority of cases. In 
Table~\ref{PhysicalParametersTable}, errors in the luminosities are
also given, which result from propagating the uncertainties in the 
IRAS fluxes, in the distance and in the luminosity correction factor.  

In this work, we also make use of the MSX PSC (version 2.3,
\citealp{Egan03}), which traces relatively warm dust, with better 
spatial resolution than IRAS. The beam size of MSX is
$\sim$\,20\arcsec, which yields a lower source confusion than IRAS. 
To characterise the MSX spectrum of the \uchii{}s in our sample, we 
made a black-body fit to the MSX bands at 14.65~\micron\ and 21.34~\micron. 
The reason for not including in the fit the MSX bands centred at 
8.28~\micron\ and 12.13~\micron\ is that they are highly dominated by 
polycyclic aromatic hydrocarbon (PAH) emission bands as well as 
the silicate feature at 9.7~\micron\ (see
\citealt{Peeters02}). Inspection of the ISO 
spectra of \uchii{}s presented in \citet{Peeters02} shows that a 
black-body fit, even in the relatively featureless range between 14 
and 21~\micron, is a poor representation of the general shape of the
spectral energy distribution. The luminosity associated to each MSX
source (\lmsx) is listed in column~(11) of 
Table~\ref{PhysicalParametersTable}. The quoted errors result from the
propagation of the errors in the black-body's temperature and scaling
factor (given by the fitting routine) and the errors in the 
distances.

\section[RESULTS]{RESULTS}
\label{Results}

We now focus on the near-IR morphology of each \uchii\ region
individually (Sec.~\ref{Results:IndividualRegions}).  
The stellar population is also discussed, particularly the most likely 
candidates for ionizing stars. The luminosity and Lyman photon rate of
these near-IR stars is compared with those inferred from IRAS and radio 
continuum data. More emphasis is given to the description of the 
regions whose near-IR sub-arcsecond morphology has not been deeply 
studied before  (\gton, \gtfo). Sources for which detailed 
adaptive-optics near-IR data are already published are discussed 
less deeply (e.g. \gf, \citealp{feldt:99}; \goo, 
\citealp{henning:2001}). In Sec.~\ref{Results:Discussion}, we
analyse the general properties of the sample. 

\subsection[Discussion on Individual Regions]{Discussion on Individual Regions}
\label{Results:IndividualRegions}

Even though in Table~\ref{PhysicalParametersTable} the stellar model 
grids by \citet{Vacca96} and \citet{Smith02} have been included
for comparison, in the following discussion on each individual object, only 
the grid by \citet{Smith02} is considered to determine spectral types 
from radio and IRAS data, since it is based in more realistic models 
than the grid by \citet{Vacca96}. In Sec~\ref{ComparisonWithRadioSec}, 
we discuss in more detail the advantages of one grid with respect to 
the other.

\subsubsection[G309.92+0.48]{\gton}
\label{gtonSubSubSec}

This region was catalogued as an \uchii\ region according to its IRAS
colours \citep{Bronfman96}. It was classified as unresolved by
\citet{Walsh98} based on their radio-continuum observations. 
\gton\ is known to be associated with \water, 
OH and methanol maser emission \citep{Braz83,Walsh99}.  

\begin{figure*}
\epsscale{2.1}
\plotone{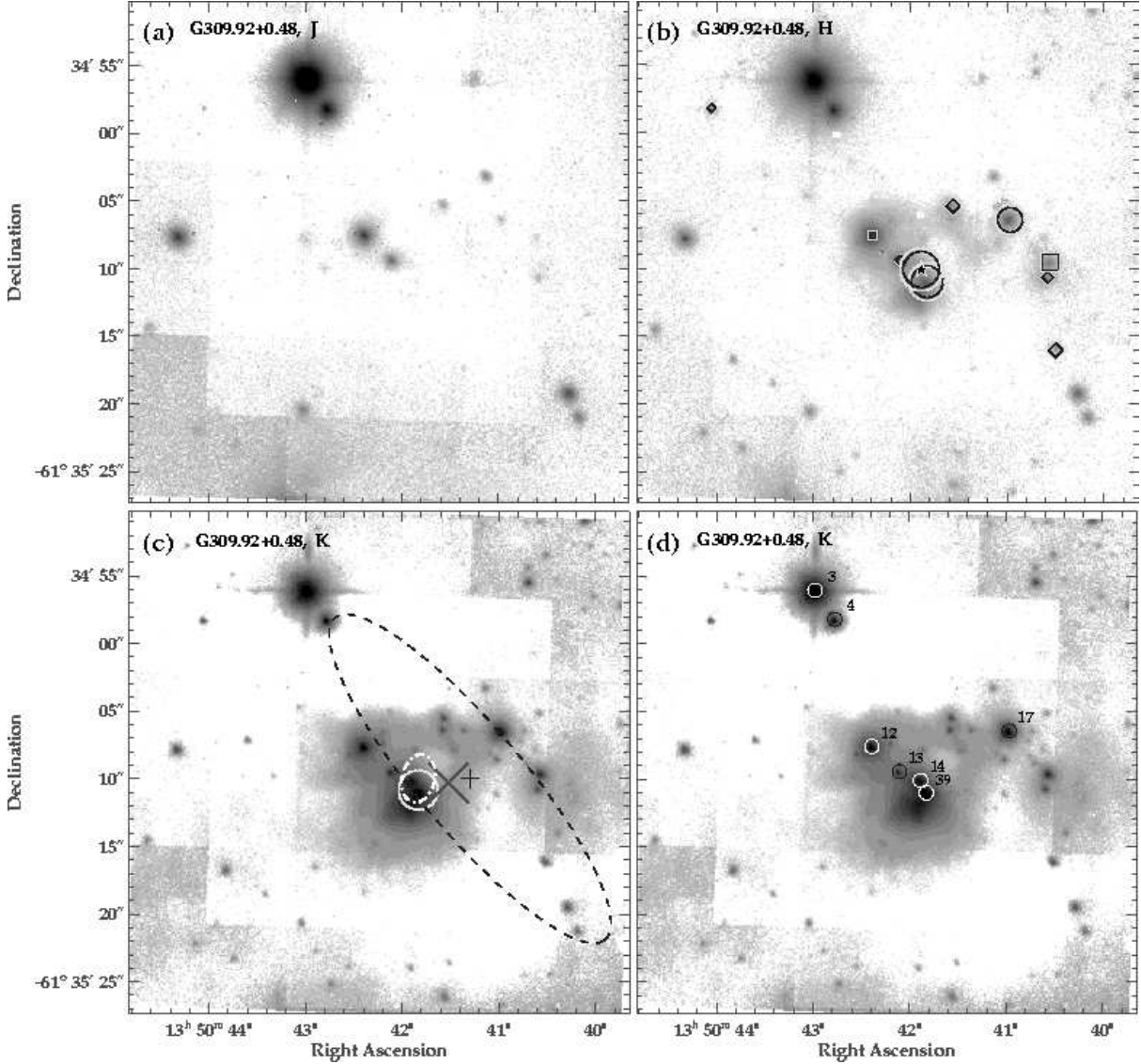}
\caption{{\scriptsize Near-IR images of G309.92+0.48. 
	{\bf (a)} J-band image. The grey-scale varies from 12.0 
	\mac\ (dark) to 20.5 \mac\ (white). The sharp-edged grey
	features in the background are artifacts from the mosaics and 
	are shown to illustrate the noise level in the images.  
	{\bf (b)} $H$-band image. The grey-scale varies from 9.0 to 
	19.5 \mac. The symbols represent stars whose \jh\ and \hks\ 
	colours are above certain cut-off values (see 
	Fig.~\ref{G309CCandCMLabel} and explanation in 
	Sec.~\ref{MassfunctionSec}). Stars with spectral types earlier 
	than O3V are represented by open circles. Spectral types in
	the range O9V\,-\,O3V are represented by open squares. 
	B9V\,-\,B0V types are represented by diamonds and A9V\,-\,A0V 
	are represented by triangles (none appears in this figure). 
	Pentagon-labelled stars (none appears in this figure) 
	have spectral types
	later than A9V. The size of the symbols is proportional to the
	\hks\ colour. Stars of any spectral type with intrinsic
	near-IR excess are labelled with a dark five-pointed star
	of fixed size. {\bf (c)} $K$-band image. The 
	grey-scale varies from 10.0 to 20.5 \mac. The solid circle
	represents the radio emission at 3.6 cm from \citep{Walsh98}.
	The plus symbol and the dashed ellipse represent the position
	and 1$\sigma$\ positional error ellipse of the IRAS source. 
	The MSX positional error ellipse (3$\sigma$) is plotted as a 
	thick grey dot-dashed ellipse. The large thick cross indicates the
	position of the
	CS core from \citet{Bronfman96}. The cross size indicates the
	pointing error of the CS peak.  
	{\bf (d)} Some of the sources selected to produce the colour-colour and
	colour-magnitude diagrams of Figure~\ref{G309CCandCMLabel} 
	are labelled on the $K$-band image. In these and the following images 
	shown in this paper, north is up and east is left. 
	\label{G309NearIRImLabel}}}
\end{figure*}

Low-resolution NIR photometry of \gton\ was obtained by \citet{Epchtein81}
and \citet{EpchteinLepine81}. More recently, the region was imaged in the 
near-IR at the seeing limit by \citet{Walsh99} and at sub-arcsecond
resolution by \citet{Henning02}. 
Figure~\ref{G309NearIRImLabel} shows our new near-IR images of 
this source taken with ADONIS at a resolution of $\sim\,0\farcs2$. 
An extended diffuse near-IR nebula around the radio 
peak is seen prominently in the $K$-band image. The nebular core,
which is located inside the 3.5~cm emitting region detected by 
\citet{Walsh98} (white solid-line circle in Figure~\ref{G309NearIRImLabel}c)
is predominantly extended towards the SE. Our images show 
that the centre of the 3.6 cm emission coincides with the position 
of source~\#39 (see
Figure~\ref{G309NearIRImLabel}) within the errors. The astrometric 
accuracy of our images is 0\farcs1, while \citet{Walsh98} quote 
an accuracy of 0\farcs05 in the peak position of
the 3.5~cm emission. In the $K$-band image shown by \citet{Walsh99}, the
radio peak appears slightly displaced ($\sim1''$) towards the south 
of the apparent position of source~\#39. We also note that the magnitudes 
listed in Table~1 of Walsh et al. are systematically shifted by
$\sim$\,-3 mag with respect to our values for common unresolved
stars. Such an increase in brightness, would shift most of the stars 
in the C-M diagram to a zone where they would appear over-luminous (at 
the assumed distance). We discard the possibility of variability in 
the sources, since the magnitude difference appears to be roughly the 
same in all sources. We therefore adopt our own values in the
following discussion.

\clearpage

\begin{figure*}
\epsscale{2.1}
\plotone{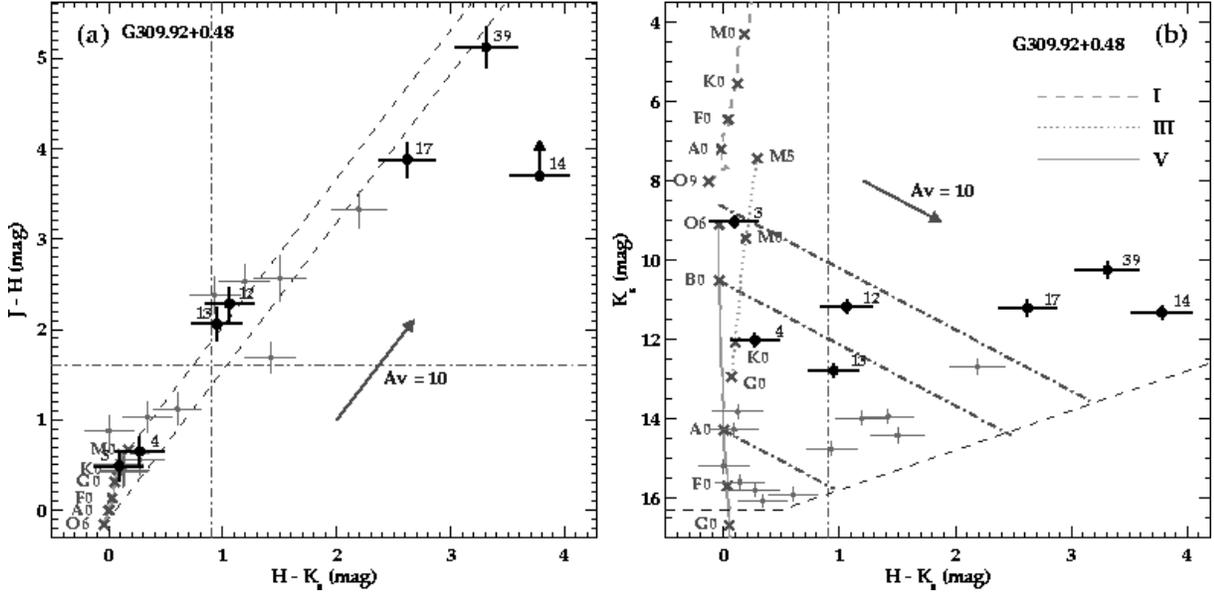}
\caption{{\bf (a)} Colour-colour diagram of G309.92+0.48. The grey solid
	line at the bottom left of the diagram represents the Main 
	Sequence, where some spectral types 
	are marked with crosses. The dashed lines show the reddened
	Main Sequence. The arrow represents a visual extinction of 10
	mag. The colours of most of the sources in our images 
	are represented by grey error bars. The colours of some 
	selected sources (listed in Table~\ref{PhotometryTable} and
	labelled in Fig.~\ref{G309NearIRImLabel}) are represented by black
	circles with error bars. Some of these sources (\#12, \#13,
	  \#14, \#17 and \#39) have been selected because they are
	  bright in the K band, they are located near the radio peak, 
	  and have considerable near-IR excess, which makes them the 
	  best candidates for ionizing sources of the \uchii\ region. 
	  The remaining sources (\#3 and \#4) are bright in the 3
	  bands. They are normally bright nearby stars with
	  hardly any extinction, and therefore, they should appear at
	  the bottom left of the C-C diagram. Hence, they were 
	  selected to check for consistency in the photometry. 
	  These selection criteria for labelling are followed in 
	  all the remaining \uchii s shown in this paper. The arrow 
	for source~\#14
	indicates that it was not detected in the $J$\ filter. The
	vertical and horizontal
	dot-dashed lines represent the \hks\ and \jh\ colour cut-offs 
	between foreground stars and stars likely belonging to the
	\uchii\ region (see discussion in Sec.~\ref{MassfunctionSec}). 
        {\bf (b)} Colour-magnitude diagram of G309.92+0.48 at a
	distance of 5.5~kpc. The unreddened Main Sequence is 
	represented by the grey solid line. The dotted line is the
	giant branch and the grey-dashed line represents the 
	super-giant branch. Spectral types are indicated by 
	crosses along each branch. A reddening vector of 10 visual 
	magnitudes is 
	represented by a grey arrow. The dark dashed line is our 
	detection limit. The colours of several sources measured 
	in our near-IR images are represented with the same symbols
	as in panel (a). The vertical dot-dashed line represents the 
	same \hks\ colour cut-off as in panel (a). Dot-dashed lines
	parallel to the extinction arrow separate different spectral 
	types. The upper of these lines corresponds to an spectral
	type O3V, whose \avm\ is 0.5 mag brighter than that of an O6V 
	star \citep{Vacca96} and we assume to have the same 
	intrinsic colours as an O6V star.
     	\label{G309CCandCMLabel}}
\end{figure*}

\clearpage

Figure~\ref{G309CCandCMLabel} shows the colour-colour and
colour-magnitude diagram of selected stars within the field of 
view of our near-IR images. We have labelled with numbers 
  the some of the brightest stars in the K band. In particular, 
  we have labelled those sources which are located near the \uchii\
  region, and which have considerable near-IR excess, since they are
  the best candidates for ionizing sources of the \uchii\ region.
Figure~\ref{G309CCandCMLabel}a indicates
that two of the most obscured sources (\#39 with \avi$\sim$60~mag, and
\#14 with \avi$>$50~mag) are located near the peak of the 3.5~cm
emission. Source~\#39 lies within the extincted Main Sequence, while
source~\#14 appears to show some IR excess. Source~\#17, which is located
at 8$''$\ (i.e. 0.2~pc at the adopted distance of 5.5~kpc) NE of the
\uchii\ region, is also highly extincted (\avi$\sim$45 mag). The extinction
appears to be lower towards the NW of the \uchii\ region, since sources~\#12
and~\#13 have visual extinctions of $\sim$\,25~mag. 

The C-M diagram 
(Fig.~\ref{G309CCandCMLabel}b) is helpful to find out which of these 
sources are most likely to be responsible for the ionization of the 
\uchii\ region. The C-M diagram indicates a spectral type either
earlier than O6V or approximately K0I for source~\#39. However, if 
source~\#39 is a super-giant, its spectral type would be too cold
(K0I) to ionize the \hii\ region, although it would contribute 
considerably to the total luminosity of the system. For source~\#14, a
somewhat more uncertain spectral type $\sim$\,O9I or earlier than O6V 
is estimated once its \hks\ excess is removed from the C-M diagram,
following the procedure described in 
Sec.~\ref{Physics:IonizingSources:NIRPhotometry}.
Source \#14 could be a super-giant and still ionize the \hii\ region. 
Source~\#17 appears to be a reddened early OV or a late AI star, which
in the later case, would be too cold to be candidate for
ionizing source. The location of source~\#12 in the C-M diagram is 
consistent with an 
intermediate OV or KIII star under moderate extinction. Source~\#13 
can be interpreted as a late O/early B Main-Sequence star or 
an early KIII star. Hence, if we discard the possibility that stars 
\#12 and \#13 are giants, and \#17 and \#39 are super-giants, we find 
up to 5 possible ionizing sources for the \uchii\ region, one of which 
(\#14) could still be a super-giant.

Once the main near-IR population has been analysed, we now focus on
the radio and IRAS data for \gton. A peak flux density 
of 350 mJy/beam at 3.5~cm \citep{Walsh98}, 
a beam size of 1\farcs3 and a T$_\mathrm{e}=7500$~K were utilised to 
calculate the beam temperature, optical depth, and 
emission measure, which yielded a log(\nlyc)=48.0 at a distance of 
5.5~kpc. The stellar models from \citet{Smith02} indicate a O9V 
spectral type for the ionizing source capable of producing such a
Lyman photon rate. Any of 
the possible ionizing sources listed in the previous paragraph would
suffice to produce such a Lyman photon rate. A total 
luminosity of $3.2\pm0.2\times10^5$~\lsun\ was inferred from the fluxes of 
IRAS\,13471-6120 given in the IRAS point source catalogue, which
indicates an O5V spectral type for the heating source. Source \#39
alone is already more luminous than the IRAS source. An anisotropic
dust distribution in the \uchii\ region or the presence of undetected 
(obscured) stars may account for this discrepancy between the IRAS
luminosity and the luminosity of the observed near-IR stellar 
population (see the general discussion in Sec.~\ref{ComparisonWithIRASSec}).  

\subsubsection[G351.16+0.70]{\gtfo (NGC\,6334-V)}
\label{gtfoSubSubSec}

\objectname[]{NGC\,6334-V} is a far-IR source \citep{McBreen79,Loughran86}
in the complex star forming region NGC\,6334. In the original
low-resolution map at 69~\micron\ by \citet{McBreen79}, NGC\,6334-V 
appears as a point source at a resolution of $\sim$3$'$. The peak of
NGC\,6334-V is located 30$''$\ towards the north of 
\objectname[]{IRAS\,17165-3554}. Radio-interferometric studies
\citep{Simon85,Rengarajan96,Walsh98,Jackson99,Argon00} revealed
several compact (radii $<$ 2$''$) continuum sources, which are located  
25$''$\ towards the south/south-east of the far-IR peak. Using the 
VLA in C configuration,
\citet{Jackson99} found that several of these compact sources are 
included in the southern rim of a large radio-shell of $\sim 1'$\ radius. 
The geometrical centre of the shell is offset $\sim 1'$ to the NE of
NGC\,6334-V. The shell shows a clumpy morphology, which leaves the
question open of whether the compact continuum sources are clumps
in the shell or independent sites of star formation. Some of these
compact radio sources are associated with water, OH \citep{Braz83,Argon00}
and methanol \citep{Walsh98} maser emission. A large-scale bipolar molecular 
outflow is also seen towards this star-forming region \citep{Phillips91}. 

NGC\,6334-V has been extensively studied at near- and mid-IR wavelengths 
\citep{HarveyGatley83,HarveyWilking84,Simon85,Straw89,Burton00}. Up to
now, the highest spatial resolution in the near-IR ($\sim$3\farcs5)
was achieved by \citet{Straw89} and \citet{Burton00}.  

\begin{figure*}
\epsscale{2.1}
\plotone{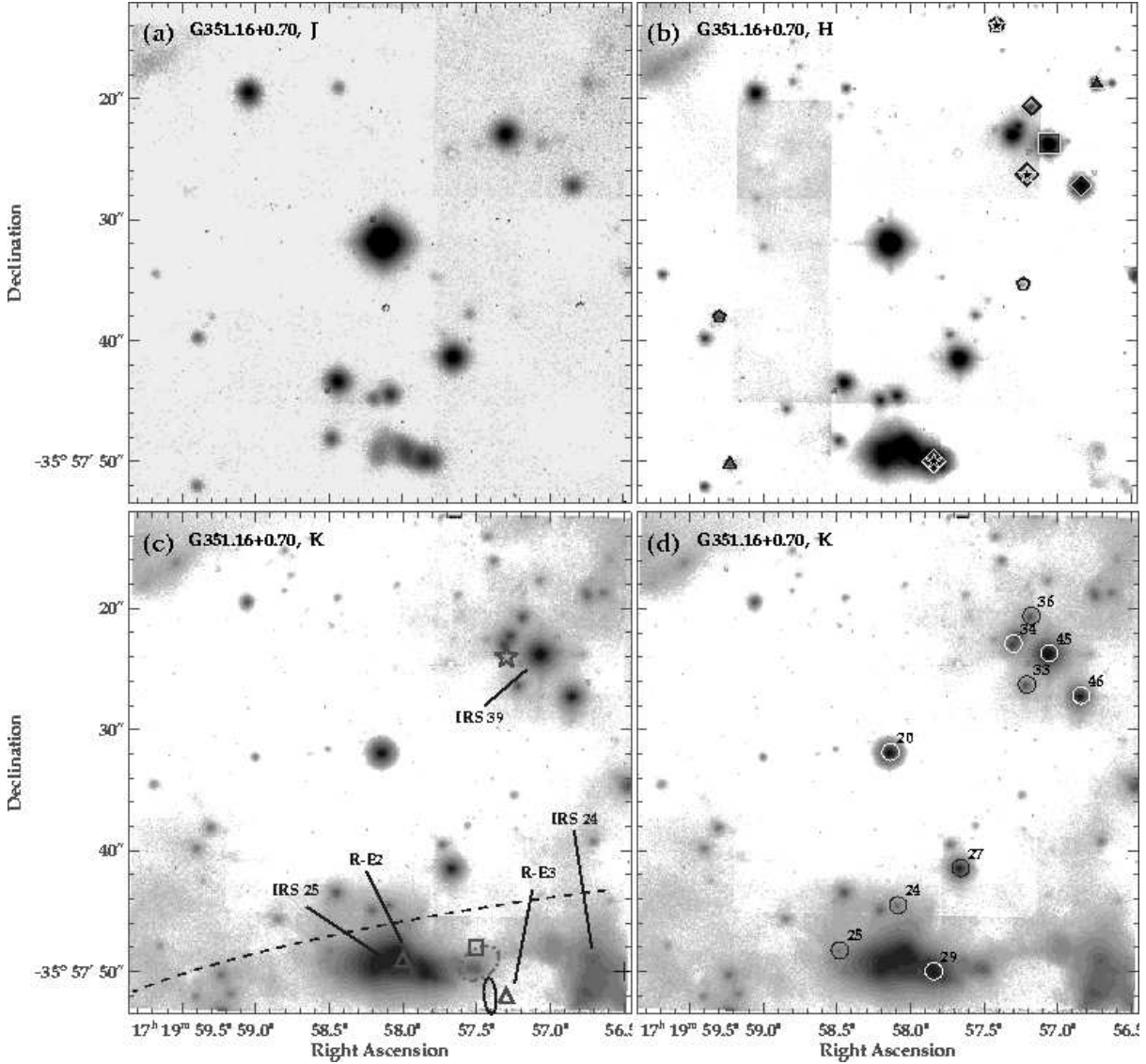}
\caption{\scriptsize AO-assisted near-IR images of G351.16+0.70/NGC\,6334-V 
        taken with ADONIS at the 3.6m telescope in La Silla. 
	{\bf (a)} $J$-band image. The greyscale varies from 14.6 
	\mac\ (dark) to 20.3 \mac\ (white). {\bf (b)} $H$-band image. The
	greyscale varies from 13.5 to 18.4 \mac. See 
	Fig.~\ref{G309NearIRImLabel} for a key to the symbols. 
	{\bf (c)} $K$-band image. The 
	greyscale varies from 9.8 to 18.9 \mac. The small solid
	ellipse represents the 3.5~cm continuum emission from 
	\citet{Argon00}. The triangles indicate the
	location of the radio-continuum sources R-E2 and R-E3
	from \citet{Rengarajan96} \citep[see also ][]{Jackson99}. 
	The plus symbol and the dashed line represent the position 
	and positional error ellipse of the IRAS source. The 
	five-pointed star represents the central
	position of the far-IR source NGC\,6334-V from \citet{McBreen79}.
	The open square indicates the location of the 20~\micron\ source
	from \citet{HarveyWilking84}. The near-IR sources IRS\,24, 25
        and 39 from \citet{Straw89} are also indicated. The MSX source
        is represented by the dot-dashed small ellipse. No CS clump 
	appears in the field of view. {\bf (d)} Some 
	of the sources selected to generate the colour-colour and 
	colour-magnitude diagrams of Figure~\ref{G351CCandCMLabel} 
	are overlaid on the $K$-band image. 
	\label{G351NearIRImLabel}}
\end{figure*}

In Figure~\ref{G351NearIRImLabel}, we present our AO-assisted images from
ADONIS in the $J$, $H$\ and $K$-bands, which have a resolution $\sim$25
times better than previous studies. In the field of view of our 
images, some of the unresolved sources from \citet{Straw89} (labelled 
IRS\,24, IRS\,25, IRS\,39) appear clearly resolved into several 
components. The bipolar reflection nebula studied by \citet{HarveyWilking84}
and \citet{Simon85} can be seen towards the south-west in our images. 
Both lobes (IRS\,24 and IRS\,25) are clearly separated from each other
and composed of several knots. Both nebular lobes are located
inside the positional error ellipse of IRAS\,17165-3554. 

No near-IR 
counterpart is detected at the position of the compact radio source
detected by \citet{Argon00} (solid ellipse in Fig.~\ref{G351NearIRImLabel}c),
which is located between both lobes of the nebula. This compact source
is also the radio source R-E3 detected by \citet{Rengarajan96} (lower right 
triangle in Fig.~\ref{G351NearIRImLabel}c). The 20~\micron\ source from
\citet{HarveyWilking84} (open square in Fig.~\ref{G351NearIRImLabel}c) 
has no near-IR counterpart either. The eastern lobe of the nebula
coincides with the position of the radio continuum source R-E2
\citep{Rengarajan96,Jackson99}. The irregular \uchii\ region detected by
\cite{Walsh98}, with coordinates (J2000) 
$\alpha$=17$^\mathrm{h}$19$^\mathrm{m}$59\fs9, 
$\delta$=-35$^\circ$57$'$40$''$, is off the field of view of our images. 
The most interesting feature in our images is the group of $\la 10$\ 
unresolved near-IR sources at the position of the source 
IRS\,39 \citep{Straw89}, which is coincident with the far-IR
peak, NGC\,6334-V.

\begin{figure*}
\epsscale{2.1}
\plotone{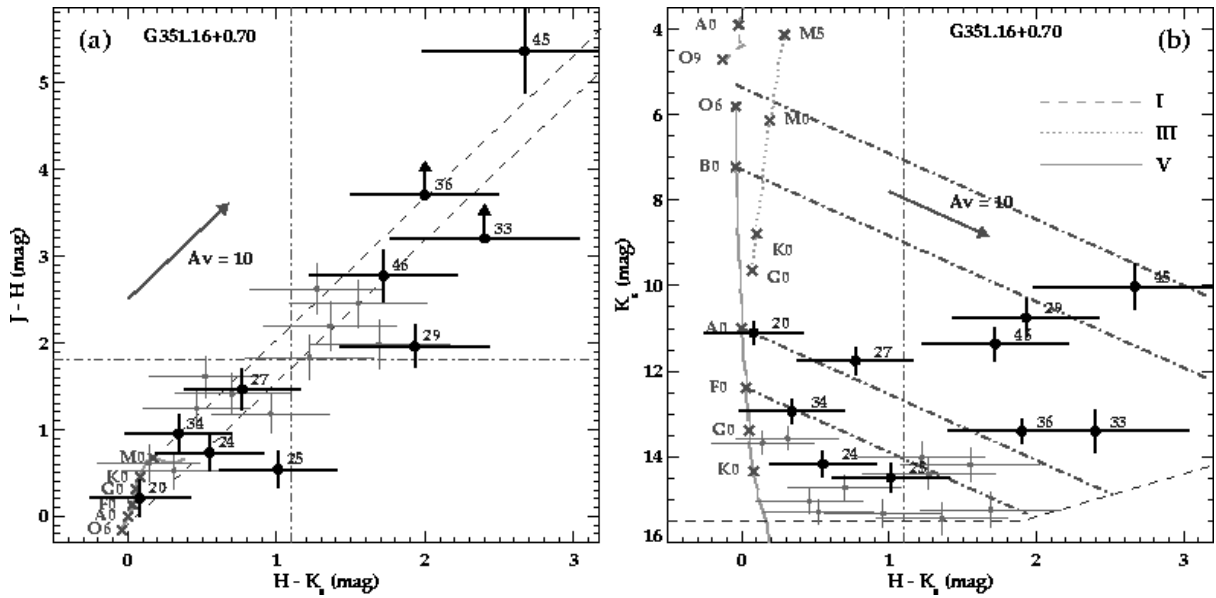}
\caption{{\bf (a)} Colour-colour and {\bf (b)} colour-magnitude
        diagram for G351.16+0.70. Key to symbols and lines is given 
	in Figure~\ref{G309CCandCMLabel}. 
	\label{G351CCandCMLabel}}
\end{figure*}

Our C-C diagram (Fig.~\ref{G351CCandCMLabel}a) indicates a reddening 
of 30 - 40 magnitudes in the visual at the location of NGC\,6334-V. 
Source~\#45 (see Fig.~\ref{G351NearIRImLabel}d) is obscured by an 
\avi\ of $\sim$40 mag. Sources~\#33 and \#36, located within 5$''$\ of 
\#45, follow in decreasing degree of obscuration, with visual extinctions
$>$30~mag each. In the same association of unresolved IR sources, 
\#46 appears to have a visual extinction of $\sim$20~mag. The
extinction near the compact region R-E3 appears to be higher, since 
no near-IR point source is detected. Source~\#29, located at $3''$\ to
the east of R-E2, apparently has an extinction of $\sim$20~mag in the visual. 
However, this source is clearly extended in our $K$-band image, probably
a knot in the nebula IRS\,25, and hence the extinction determination
is very uncertain.

The C-M diagram shown in Figure~\ref{G351CCandCMLabel}b yields
further insights into the stellar population. Associated with the 
far-IR source NGC\,6334-V, we identify one O6V (\#45) star, and 
three BV stars (\#33, \#36 and \#46). Note that the de-reddened colours 
of these four point sources are also consistent with spectral types in 
the early-to-mid MIII. In principle, this possibility cannot be 
discarded, except for the presumed youth of NGC\,6334-V. In the same 
association, \#34 appears to be barely extincted,
indicating that this is likely a foreground star. In the inmediate 
surroundings of the R-E3 region no near-IR point source was detected,
probably due to high extinction. This is supported by the presence of
an MSX source located at the same position as RE-3. 

\clearpage

\clearpage

The large-scale shell, with a radius $\sim$0.3~pc, detected by 
\citet{Jackson99} 
at 3.5~cm, has an integrated flux density of 1.2~Jy. At a distance of 
1.2~kpc, a log(\nlyc)=47.3 is needed to ionize the shell with a 
T$_\mathrm{e}=7500$~K. If we
compare this value with the predictions from the stellar models of 
\citet{Smith02}, the inferred spectral type for the ionizing star of
the large scale shell is B0V. If the radio peaks R-E2 and
R-E3 are considered to be independent \uchii{}s rather than clumps 
(with no stars inside) forming
part of the large-scale shell, similar lower limits for the spectral
types are inferred from their 3.5~cm flux 
(1.4 - 1.7~mJy; \citealt{Jackson99}). No near-IR point-like 
counterpart is found coincident with any of the compact radio sources. 

We focus our attention on the compact association of near-IR point
sources resolved at the position of IRS\,39. We 
infer a spectral type for source~\#45 (O6V) which would produce more 
than enough Lyman photons to ionize the large-scale shell. Besides, 
sources~\#33, \#36 and~\#46 (early BV spectral types) can also 
play a role in the ionization of the shell. However, the fact that 
this stellar association is located near the southern rim of the
shell, rather than near the centre makes it quite unlikely that 
they are the main ionizing sources for the whole shell. If source~\#45
is contributing partially to the ionization of the southern rim of
the shell, it is difficult to explain why no traces of ionized gas are
seen at the position of source~\#45 itself. The possibility that 
\#45, \#33, \#36 and \#46 are protostars can be discarded, since they 
show low intrinsic $K$-band excess in our C-C diagram. 


Two possible spectral types can be determined for the dust heating
sources from the mid- and far-IR data available for \gtfo. The first 
estimate is based on the fluxes from \hbox{IRAS\,17165-3554}. A total
luminosity of $4\pm1\times10^4$~\lsun\ yields a spectral type for
the heating source later than B0V (see
Table~\ref{PhysicalParametersTable}). 
The second estimate is obtained by
scaling the total luminosity given in \citet{Loughran86} for the
far-IR source NGC\,6334-V ($1.7\times10^5$~\lsun\ at 1.7~kpc) to our 
assumed distance of 1.2~kpc. We obtain a new \ltot=$8.5\times10^4$~\lsun,
which yields an O8V spectral type for the heating source. 

The IRAS source is associated with the western lobe of the bipolar nebula
(IRS\,24). Since this is at the edge of the field of view of our
images, we do not have photometric information on the stellar
population to compare with the IRAS luminosity. The situation is
different in the case of the far-IR source NGC\,6334-V, which is
clearly associated with the group of embedded near-IR sources \#45,
\#33, \#36 and \#46 in our images. The spectral type inferred for the
near-IR sources (one $<$O6V and three early-to-mid BV) is clearly
earlier than the O9.5V needed to explain the total luminosity associated
with NGC\,6334-V. Therefore, these near-IR sources maybe contributing
only partially to the heating of NGC\,6334-V.

\subsubsection[G5.89-0.40]{\gf}
\label{gfSubSubSec}

\gf\ was classified by \citet{wc89} as a shell-type \uchii\ region of 
diameter 5$''$. \citet{KimKoo01} found that the compact radio
emission is located near the centre of a 15$'$\ extended ionized
halo. Several OH, \water, and \methanol\ maser spots have been 
found in the region \citep{Argon00,Hofner96,Walsh98}. 
\gf\ is known to be associated with an outflow, whose orientation 
has been found to be different (E-W,N-S,NE-SW) depending on the author 
and the tracer \citep{Harvey88,Cesaroni91,Acord97,Sollins04}. 

Near-IR images at the seeing limit were obtained by \citet{Harvey94}. 
A detailed study of this \uchii\ region from near-IR to millimetre
wavelengths was presented by \citet{feldt:99}. They show the first 
AO-assisted near-IR images of the region, with a resolution of 
0\farcs4 in the $K$~band. 

\begin{figure*}

\epsscale{2.1}
\plotone{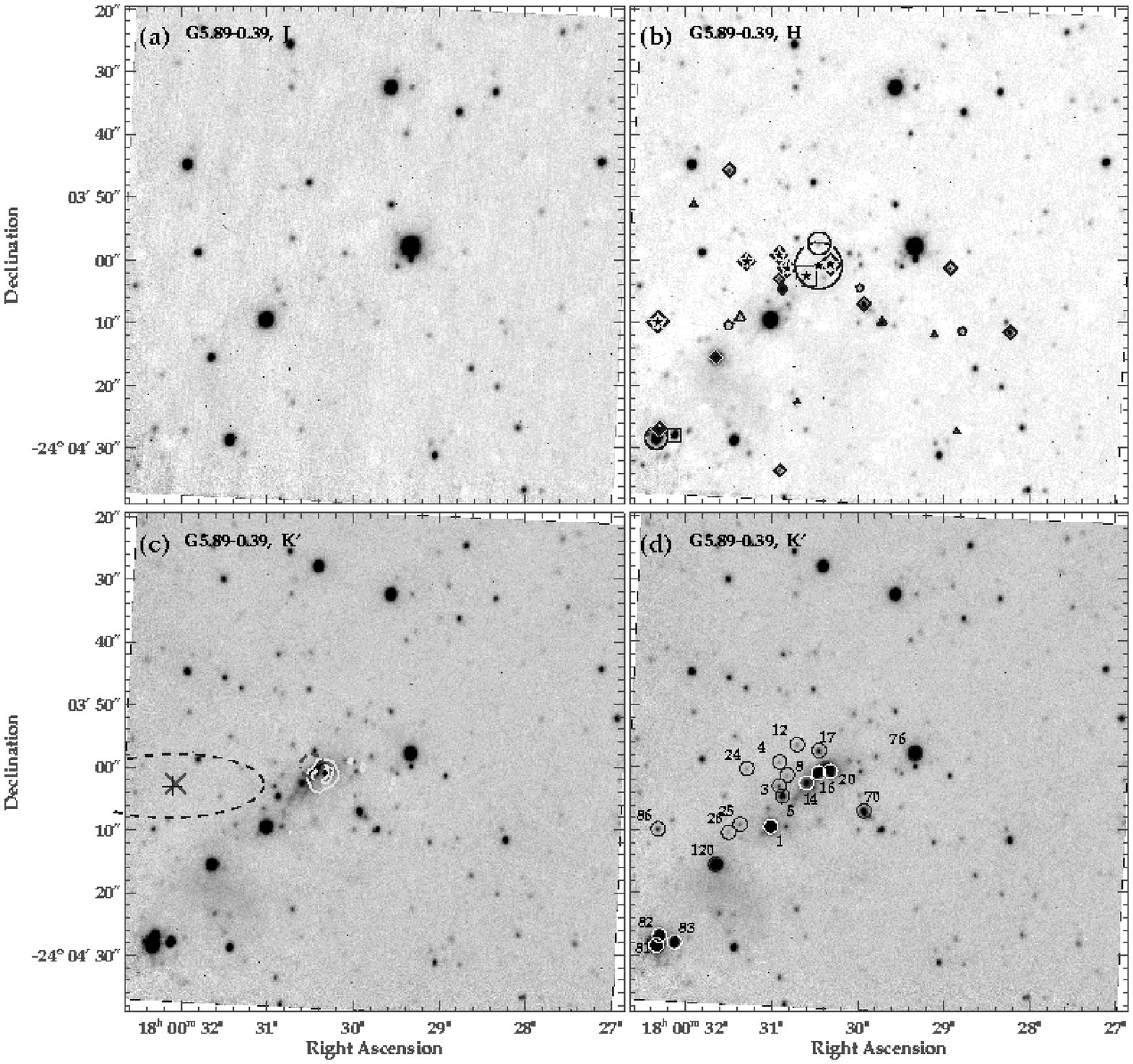}
\caption{Near-IR images of G5.89-0.39. 
	{\bf (a)} $J$-band image. The greyscale varies from 14.5 
	\mac\ (dark) to 15.0 \mac\ (white). {\bf (b)} $H$-band image. The
	greyscale varies from 12.8 to 13.0 \mac. See 
	Fig.~\ref{G309NearIRImLabel} for a key to the symbols.
	{\bf (c)} $K'$-image. The 
	greyscale varies from 10.06 to 10.14 \mac. The white contours
	represent the radio-emission at 2 cm. Contour levels are at 
	5, 15 and 25$\sigma$. Symbols are the same as in
	Fig.~\ref{G309NearIRImLabel}.
	{\bf (d)} Some of the stars selected to produce the 
	colour-colour and colour-magnitude diagrams of 
	Figure~\ref{G589CCandCMLabel} are overlaid on the $K'$-band 
	image. \label{G589NearIRImLabel}}
\end{figure*}

Our new high-resolution near-IR images of \gf\ taken with ALFA are
presented in Figure~\ref{G589NearIRImLabel}. The resolution of these
data is comparable to that shown in \citet{feldt:99} 
\citep[see also][]{Henning02}. Our images show basically the same 
features as in \citet{feldt:99}. Nevertheless, we include them here 
for completeness. 

\begin{figure*}
\epsscale{2.1}
\plotone{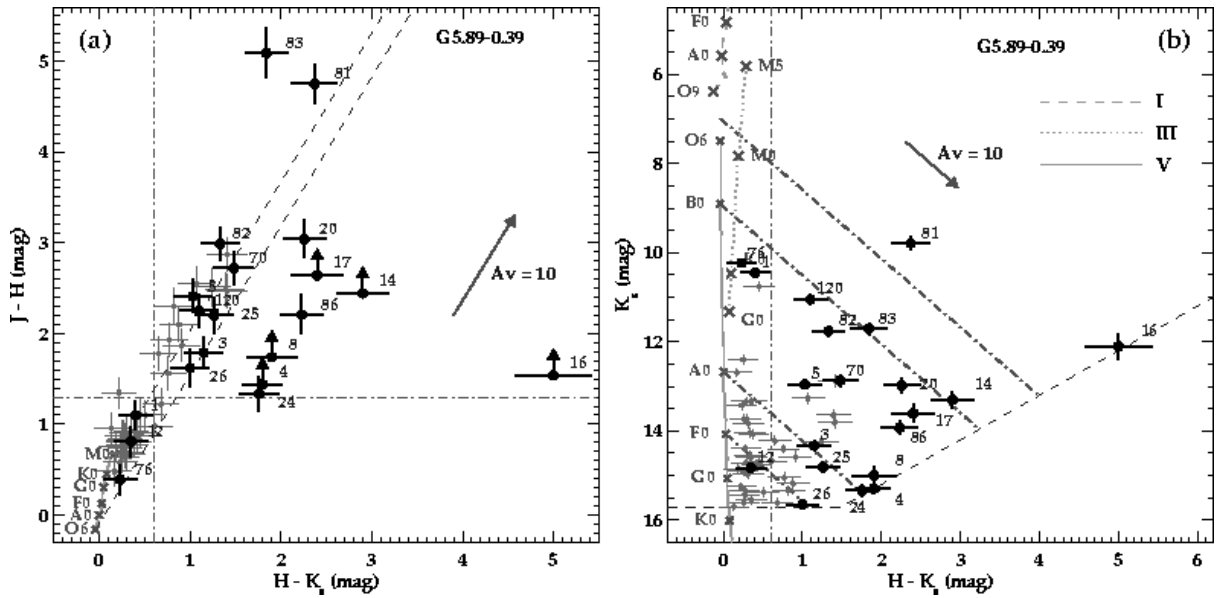}
\caption{{\bf (a)} Colour-colour and {\bf (b)} colour-magnitude
        diagram for G5.89-0.39. Key to symbols and lines is given in 
	Fig.~\ref{G309CCandCMLabel}.
	\label{G589CCandCMLabel}}
\end{figure*}

The C-C diagram of this region is shown in Figure~\ref{G589CCandCMLabel}a 
(we use the same notation as in \citealt{feldt:99}). Several sources
appear to have \hbox{\hks} excesses, although we focus our attention 
on sources~\#14, \#16 \#17 and \#20, since they are very red and are 
located within or very close to the radio \uchii\ region. Source~\#14 
appears as an unresolved core surrounded by an extended halo 
in our ALFA \ks-band image, while source~\#16 is only barely
resolved.  Source \#17 is unresolved and source~\#20 appears to be 
clearly extended. In any case, we tried to assign a spectral type to all 
of them, by using the method described in 
Sec.~\ref{Physics:IonizingSources:NIRPhotometry}. Sources~\#14 and
\#17 appear to be late BV stars under $\sim$\,20~mag of visual
extinction. The photometry of source~\#20 is consistent with a
B5V star under a \avi\,$\sim$\,25~mag. Source~\#16 shows an extremely
large excess, \hks\,$\sim$\,4~mag, which yields a $\sim$\,F5V spectral 
type, under the assumption that all the excess is in the \ks\ 
band. In the cases of sources \#14, \#17, and \#20, giant spectral 
types from KIII to GIII are also possible based only on the photometry, but 
then these stars would lack of any ionizing capabilities.

AO-assisted Fabry-Perot imaging shows that \#14 and \#20 are strong 
Br$\gamma$\ emitters, which would explain part of their IR excess 
\citep{Puga04b}. The same data indicate that part of 
the emission in \#16 is also due to Br$\gamma$. Recent $K$- and
$L'$-band imaging with NAOS/CONICA at the VLT, with higher resolution
and sensitivity that our ALFA images, indicate that source~\#16
actually contains a star of spectral type O5V \citep{Feldt03}, which 
is likely to be the ionizing star of the \uchii\ region due to its 
location within the shell. Hence, our assumption of all the \hks\ 
excess coming from the \ks-band appears to yield a far too late 
(F5V) spectral type. Therefore, we adopt hereafter the spectral type 
inferred by \citet{Feldt03}, which scaled form their adopted distance
of 1.9~kpc to our value of 2.6$\pm$0.6~kpc, is $\sim$\,O3V.

The spectral type of the ionizing star inferred from radio data is 
O8V (see Table~\ref{PhysicalParametersTable}). 
The spectral type of a single star necessary to produce the IRAS
emission is O7V. Hence, in \gf, the near-IR spectral type of the best
candidate for ionizing the \hii\ region is
earlier than the radio and IRAS spectral type. This is a common
feature of almost all \uchii{}s studied here, which will be discussed in 
Secs.~\ref{ComparisonWithRadioSec} and~\ref{ComparisonWithIRASSec}.

\clearpage

\clearpage

\subsubsection[G11.11-0.40]{\goo}
\label{gooSubSubSec}

This \uchii\ region, which was classified as irregular by \citet{Kurtz94},
has a well-defined core at 3.6~cm, with a halo that extends 10$''$\
towards the SE. It is associated with methanol maser emission
\citep{Walsh97}, and with high-velocity CO emission
\citep{Shepherd96}. 

The near-IR data of \goo\ presented here form part of the previous work
shown in \citet{henning:2001}. Our new improved photometry --\,we use a
more appropriate model for the PSF (see Sec.~\ref{sec:Photometry}) 
and calibrate the photometry using the 2MASS PSC\,--\ yields $H$\
and $K_s$\ brightnesses 1~mag fainter than in \citet{henning:2001}. 

\begin{figure*}
\epsscale{2.1}
\plotone{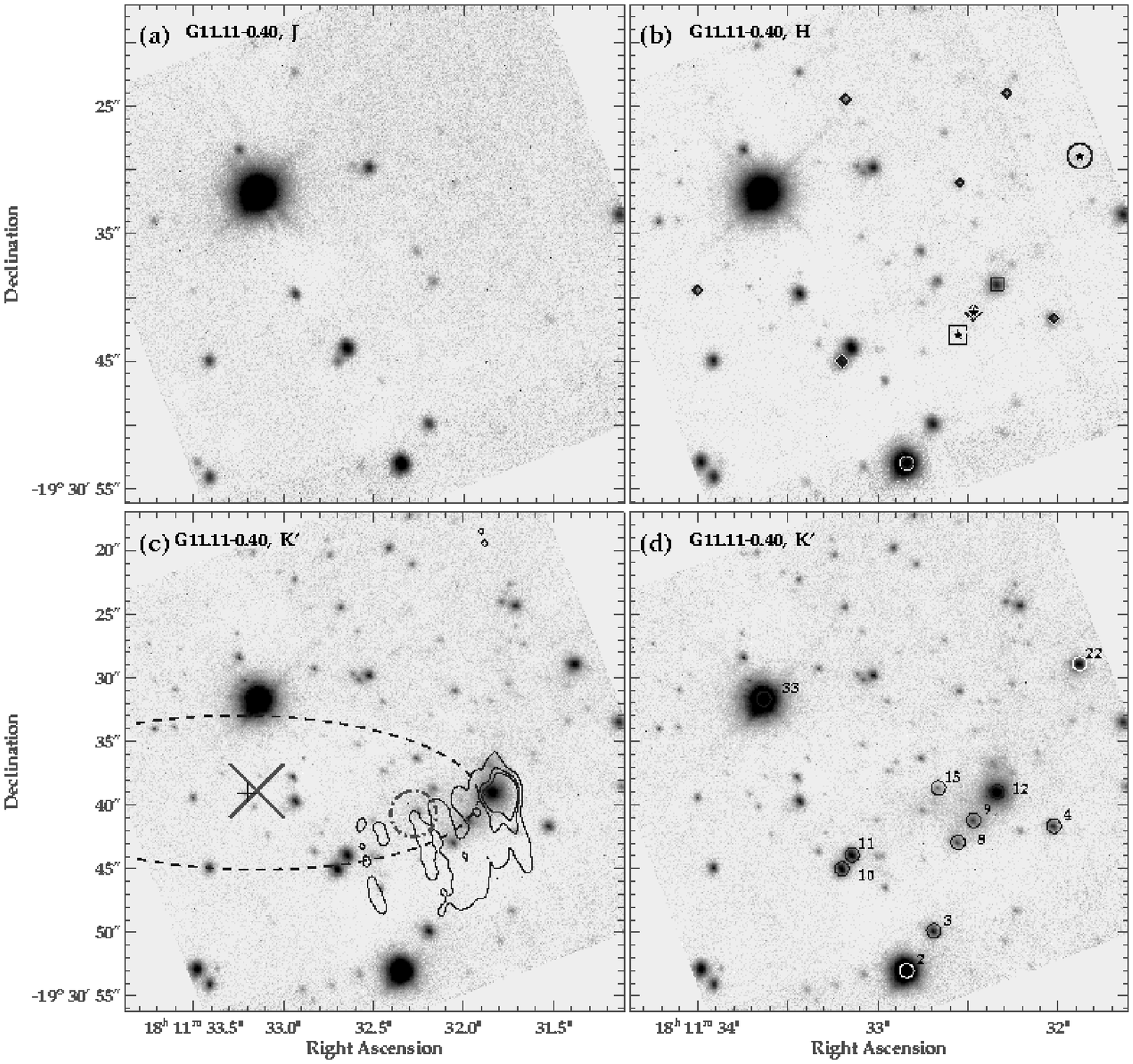}
\caption{Near-IR images of G11.11-0.40. 
	{\bf (a)} $J$-band image. The greyscale varies from 13.5 
	\mac\ (dark) to 17.7 \mac\ (white). {\bf (b)} $H$-band image. The
	greyscale varies from 13.0 to 17.9 \mac. See 
	Fig.~\ref{G309NearIRImLabel} for a key to the symbols.
	{\bf (c)} $K'$-image. The 
	greyscale varies from 12.4 to 17.2 \mac. The contours
	represent the radio-emission at 3.6 cm. Contour levels are at 
	5, 15 and 25$\sigma$. Symbols are the same as in
	Fig.~\ref{G309NearIRImLabel}. {\bf (d)} Some of the stars 
	selected to produce the colour-colour and colour-magnitude 
	diagrams of Figure~\ref{G11CCandCMLabel} are overlaid on the 
	$K'$-band image. \label{G11NearIRImLabel}}
\end{figure*}

\begin{figure*}
\epsscale{2.1}
\plotone{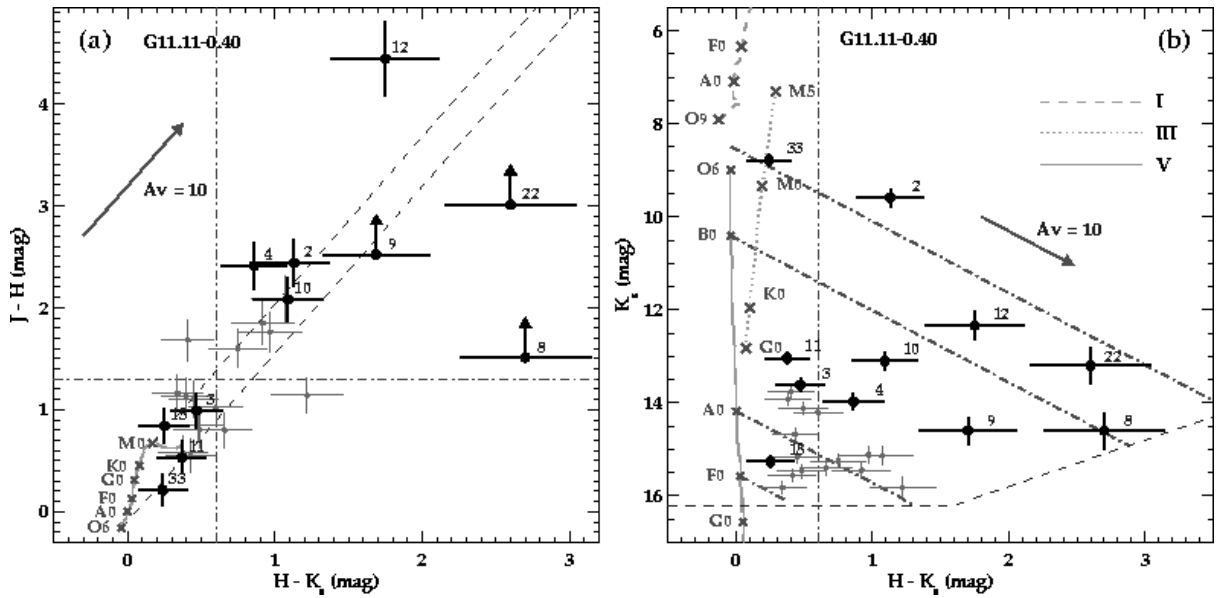}
\caption{{\bf (a)} Colour-colour and {\bf (b)} colour-magnitude 
        diagram for G11.11-0.40. Key to symbols and lines is given 
	in Figure~\ref{G309CCandCMLabel}.
	\label{G11CCandCMLabel}}
\end{figure*}

The near-IR morphology of this region is shown in
Figure~\ref{G11NearIRImLabel}. We present the C-C and C-M diagrams in
Figure~\ref{G11CCandCMLabel}. The most obscured sources within
or near the radio-continuum emission are \#8, \#9, \#12 and \#22. 
Star \#22, located some 10\arcsec\ (0.25~pc) to the NE of the 
\uchii\ region, also appears to suffer high extinction. The 
photometry of source~\#8 is consistent with an early AV star 
under a \avi\,$\sim$\,10~mag. Sources~\#9 and~\#22 are consistent with
with late and early BV spectral types, respectively, with a visual
extinction of $\sim$\,25~mag in both cases. Source \#12, which is 
the closest to the radio peak, appears to be bluer than the reddened
MS in the C-C diagram. The spectral type inferred from 
the C-M diagram is that of an O6V  with a visual extinction of 
$\ga 35$~mag. Note that in this case no correction for the
near-IR excess was made.

From the 2~cm flux density listed in \citet{Kurtz94}, we infer a
log(\nlyc)=47.8, i.e. a spectral type O9V for the
ionizing source. This spectral type is later than the 
O6.8~ZAMS obtained by \citet{henning:2001} because they used the peak 
flux density to derive the electron density, and assumed this density to be 
uniform over a sphere of radius 0.2~pc. Here, we do the calculation
using the integrated flux density within a sphere of 0.2~pc \citep{Kurtz94}.
The spectral type inferred from the IRAS fluxes is O9V, which is two
spectral sub-types later that near-IR photometric spectral type of
source \#12. This \uchii\ region represents one of the instances where the 
near-IR colours of the best candidate to be the ionizing star indicate 
a spectral type earlier than the one inferred from both radio and 
IRAS fluxes. 

\clearpage

\clearpage

\subsubsection[G18.15-0.29]{\goe}
\label{goeSubSubSec}

\begin{figure*}
\epsscale{2.1}
\plotone{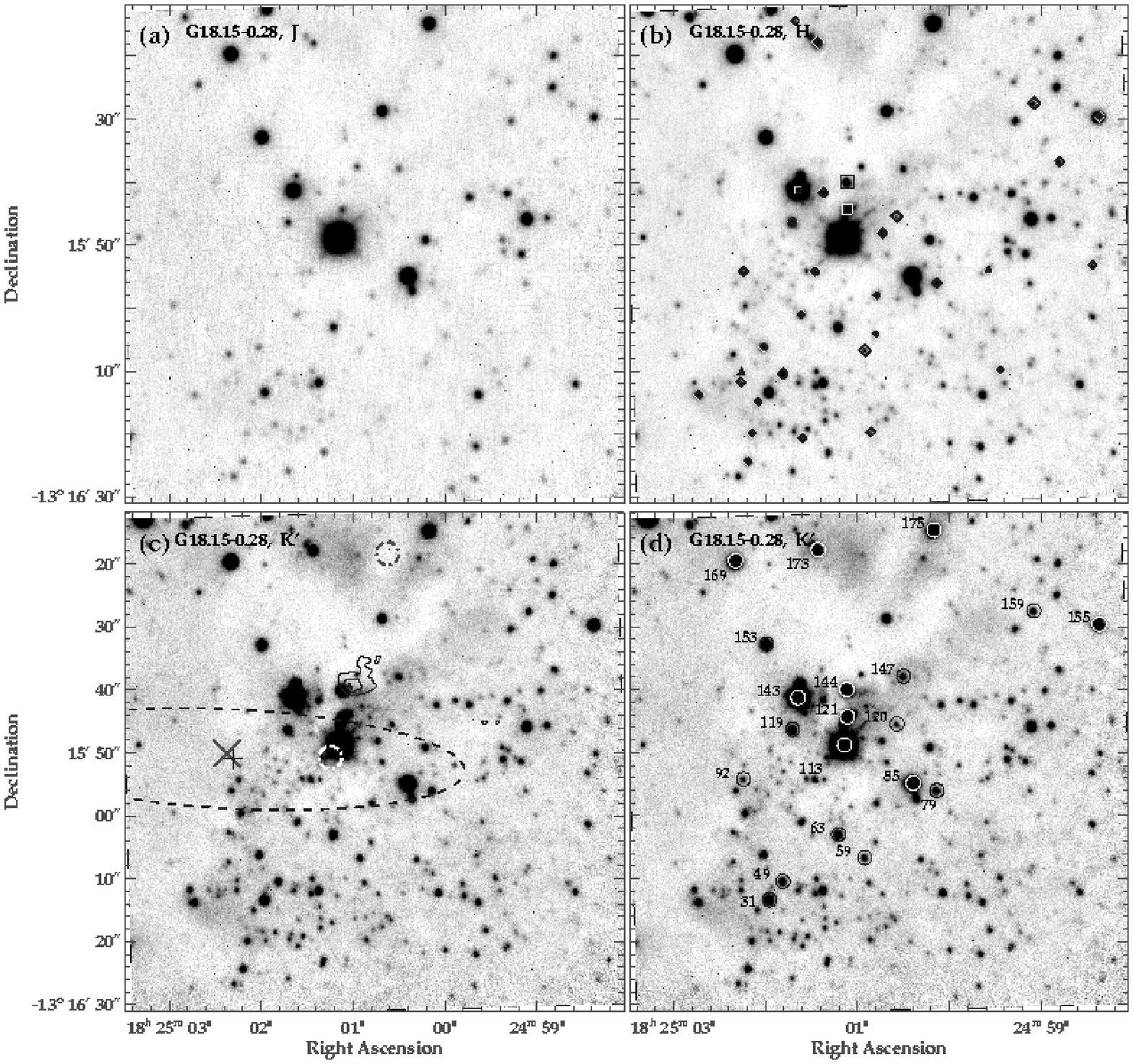}
\caption{Near-IR images of G18.15-0.29. {\bf (a)} $J$-band image. The
        greyscale varies from 14.8 \mac\ (dark) to 15.4 \mac\ 
	(white). {\bf (b)} $H$-band image. The greyscale varies from 
	13.6 to 13.9 \mac. See 
	Fig.~\ref{G309NearIRImLabel} for a key to the symbols.
	{\bf (c)} $K'$-image. The greyscale varies 
	from 11.10 to 11.14 \mac. The contours represent the
        radio-emission at 3.6 cm. Contour levels are at 
	3, 6, 9 and 12$\sigma$. Symbols are the same as in
	Fig.~\ref{G309NearIRImLabel}. In this case, two MSX sources
        appear in the field of view (clear and dark small 
        dot-dashed circles). {\bf (d)} Some of the stars selected to 
	produce the colour-colour and colour-magnitude diagrams of 
	Figure~\ref{G18CCandCMLabel} are overlaid on the $K'$-band image. 
	\label{G18NearIRImLabel}}
\end{figure*}

Little is known about this \uchii\ region classified as cometary by
\citet{wc89}. The radio source lies at the edge of a extinction
lane that covers part of the eastern region in the FOV of our near-IR 
images (see Fig.~\ref{G18NearIRImLabel}). The C-C and C-M diagrams show
that star \#144, which is located within the radio \uchii\ region, is the
most obscured source. The C-C diagram indicates reddening towards this
source due to pure extinction, with no intrinsic IR excess. The spectral 
type obtained from the C-M diagram is that of an O6V star under 30 
magnitudes of visual extinction. Sources \#143 and \#121, with
spectral types O7V and B0V and visual extinctions of 10 and 20 mag 
respectively, are also potential contributors to the ionization of 
the \hii\ region.  

\citet{Kurtz94} infer a log(\nlyc)\,=\,46.8 from the integrated flux
density at 3.6~cm, which implies an B0.5V spectral type for the
ionizing source. Therefore in this case, it appears that the
radio data underestimate by far the Lyman continuum photon rate 
associated to massive the stars detected with our near-IR photometry 
at and in the surroundings of the \uchii\ region. The IRAS fluxes yield an 
O7V spectral type, which is in relatively good agreement with our 
near-IR photometry, unless we add up the luminosities from 
the brightest sources associated with the \hii\ region (\#121, 
\#143 and \#144).   

\clearpage

\begin{figure*}
\epsscale{2.1}
\plotone{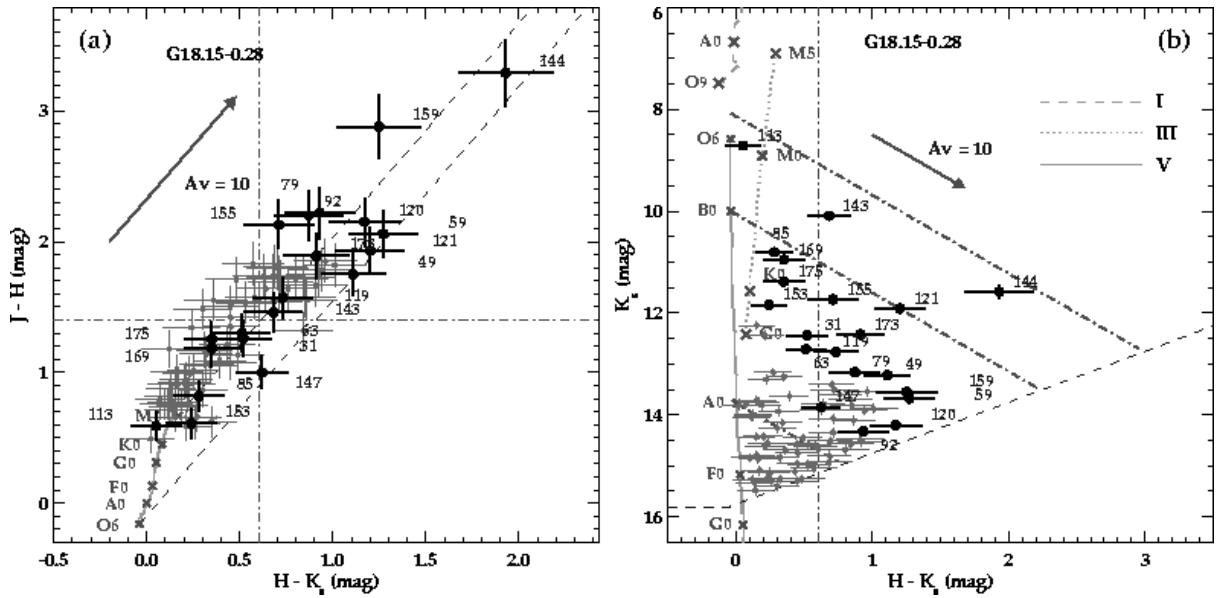}
\caption{{\bf (a)} Colour-colour and {\bf (b)} colour-magnitude 
	for G18.15-0.29. Key to symbols and lines is given in 
	Figure~\ref{G309CCandCMLabel}. 
	\label{G18CCandCMLabel}}
\end{figure*}

\clearpage

\subsubsection[G61.48+0.09]{\gso}
\label{gsoSubSubSec}

\begin{figure*}
\epsscale{2.1}
\plotone{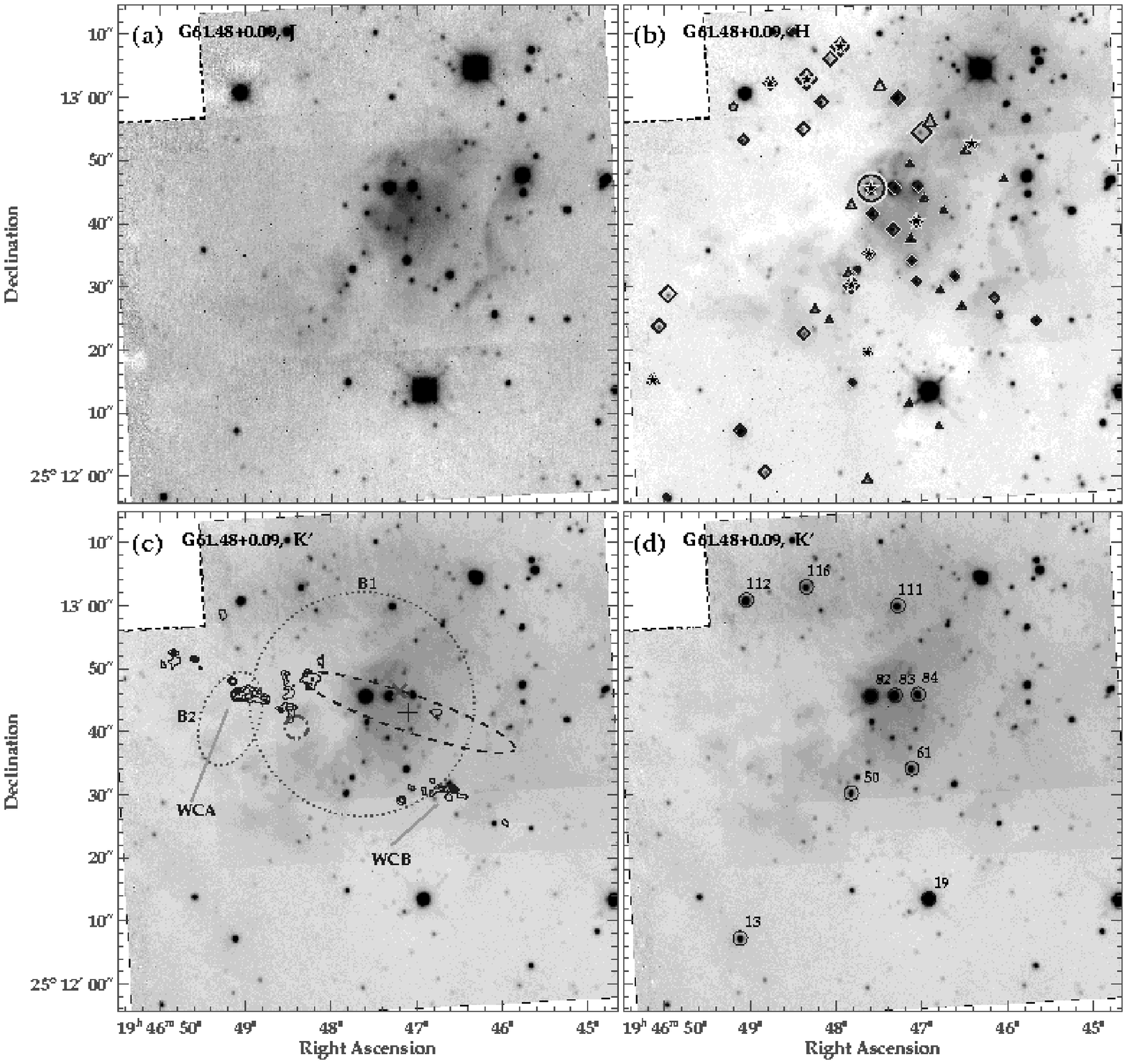}
\caption{Near-IR images of G61.48+0.09. 
	{\bf (a)} $J$-band image. The greyscale varies from 14.8 
	\mac\ (dark) to 15.2 \mac\ (white). {\bf (b)} $H$-band image. The
	greyscale varies from 13.3 to 13.6 \mac. See 
	Fig.~\ref{G309NearIRImLabel} for a key to the symbols.
	{\bf (c)} $K'$-image. The 
	greyscale varies from 10.60 to 11.72 \mac. The contours
	represent the radio-emission at 6 cm with the VLA in B
	configuration \citep{wc89}. Contour levels are at 
	3, 6, 9, 12 and 15$\sigma$. The radio sources denominated 
	G61.48+0.09A and B in \citet{wc89} are here labelled 
	as WCA and WCB, respectively. Symbols are the same as in
	Fig.~\ref{G309NearIRImLabel}. In addition, the two thin 
	dotted ellipses represent the radio components B1 and 
	B2 from \citet{Garay93}. {\bf (d)} Stars selected to produce the 
	colour-colour and colour-magnitude diagrams of 
	Figure~\ref{G61CCandCMLabel} are overlaid on the $K'$-band image. 
	\label{G61NearIRImLabel}}
\end{figure*}

\gso\ is a complex of two \uchii{}s located in the emission nebula
Sh2-88B \citep{Felli81}. Here, we focus our study on \gso{}B, which
itself has two components (see Figure~\ref{G61NearIRImLabel}). B2 is 
the eastern region, classified as spherical or unresolved. B1, the
western component, was classified as an extended cometary \hii\
region, which is undergoing a champagne flow
\citep{Garay94,Garay98a,Garay98b}. The presence of molecular gas
associated with these \hii\ regions was inferred from CO 
and CS measurements \citep{Schwartz73,Blair75}. The velocity structure
of the CO lines indicate the presence of several molecular outflows in
\gso\ \citep{Phillips91,WhiteFridlund92}.

This region has been previously studied in the near-IR by
\citet{Evans81} and \citet{Deharveng00}, the latter with a
resolution comparable to our AO-assisted images. In a separate 
paper, we \citep{Puga04a} present a detailed description 
of this region, by combining AO-assisted near-IR polarimetry, 
narrow-band imaging and radio data with part of the photometry 
from this mini-catalogue. 

In Figure~\ref{G70CCandCMLabel}, we present our C-C and C-M diagrams for
\gso. There is some slight discrepancy between our magnitudes 
(see Table~\ref{PhotometryTable}) and the
data shown in \citet{Deharveng00}, although both sets of data
agree within the errors. Stars \#82 and \#83 appear to be the main
contributors to the ionization, due to their high luminosity,
proximity to the radio peak and degree of obscuration. From our 
photometry, we infer an O9I spectral type for \#82 and a B0V type 
for \#83, under \avi{}'s of 35 and 15 mag, respectively. For this
estimate, we have subtracted 0.2 mag from the $K_s$\ magnitude of
source \#82 due to Br$\gamma$\ emission \citep{Puga04a}. The
super-giant nature of star~\#82 is reinforced when its $L'$\ 
magnitude is also considered \citep{Puga04a}. 

The radio data yield different values for the number of Lyman continuum
photon rate depending on which assumptions are made for the geometry of the
radio-emitting region. In the case of the B2 component, if we scale 
the value of \nlyc\ given in \citet{wc89}\footnote{Note that source
  \gso{}A in \citet{wc89} corresponds to source B2 in the notation of 
  \citet{Garay98a}, which we adopt.} (log(\nlyc)\,=\,46.2 at
2.0~kpc) to a distance of 2.7~kpc, we obtain a spectral type B1V for 
the ionizing source (see Table~\ref{PhysicalParametersTable}). For this
estimate, \citet{wc89} used the integrated flux and size at 6~cm 
measured with the VLA in B configuration. \citet{Deharveng00}, however, 
based their calculation on radio data taken at lower spatial resolution 
\citep{Felli81,Garay93}. They infer a log(\nlyc)\,=\,48.4 and 
log(\nlyc)\,=\,47.4 at a distance of 2.4~kpc for components B1 and B2,
respectively. These Lyman photon rates scaled to a distance of
2.7~kpc imply spectral types for components B1 and B2 
of $\sim$\,O8V and $\sim$\,B0V, respectively. 
Furthermore, \citet{Puga04a} show that different (and equally valid)
assumptions on the geometry of the ionized region, yield variations 
in the Lyman photon rate of up to one order of magnitude.
The spectral type inferred from the IRAS source associated with 
\gso\ is an O7.5V. 

Summing up, the spectral type for the hottest star derived from our 
near-IR photometry ($\sim$\,O9I) is one luminosity class higher than
the radio spectral type of component B1 ($\sim$\,O9V) and also
two spectral sub-types earlier than the radio spectral type of component 
B2 ($\sim$\,B1V).  An O9I star is 3.5 times more luminous than the 
luminosity inferred from the IRAS fluxes. This difference maybe
explained if a population of stars hidden in our near-IR images 
are the actual ionizing sources of this complex \hii\ region, as 
well as the heating stars of the IRAS source. Part of this population 
may have already started to show up in recent $L'$-band NACO observations  
\citep{Puga04a}. An anisotropic dust distribution, which would be
only partially heated by the detected near-IR population may also
cause this difference. 

\clearpage
    
\begin{figure*}
\epsscale{2.1}
\plotone{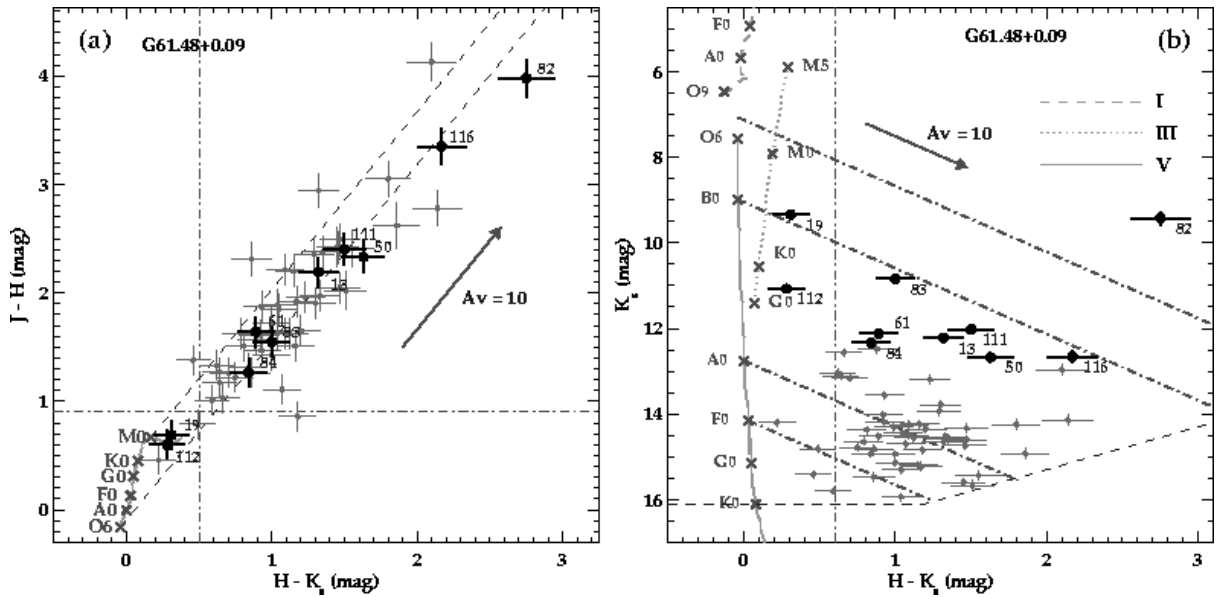}
\caption{{\bf (a)} Colour-colour and {\bf (b)} colour-magnitude 
	diagram for G61.48+0.09. Key to symbols and lines is given 
	in Fig.~\ref{G309CCandCMLabel}.
	\label{G61CCandCMLabel}}
\end{figure*}

\clearpage

\subsubsection[G70.29+1.60]{\gsvo\ (K3-50A)}
\label{gsvoSubSubSec}

\gsvo\ is a well-studied \uchii\ region, classified as a
compact radio shell by \citet{Turner84} and as a core-halo source by
\citet{Kurtz94}. \gsvo\ is the brightest and youngest of a complex of 4
radio sources (K3-50A to D) spreaded over an area of 3\farcm5
(i.e. 8~pc at the adopted distance of 8.2~kpc). \gsvo\ is coincident 
with a 10~\micron\ peak \citep{WynnWilliams77}, and also with
a CS core \citep{Bronfman96}.

Studies of radio-recombination line emission suggest the presence
of moving ionized material
(e.g. \citealt{RubinTurner69,Wink83,Roelfsema88,dePree94}). In particular,
\citet{dePree94} show that \gsvo\ is undergoing a high-velocity
bipolar outflow in the NW-SE direction. Based on CO observations,
\citet{Phillips91} infer a bipolar outflow in the NE-SW direction.

Seeing-limited near-IR imaging was performed by \citet{Howard96}. They
obtained an extinction map of the region as well as reinforced the
idea of an outflow of ionized gas with a roughly north-south
orientation. \citet{Okamoto03} did a detailed study of the stellar
population and ionization structure in the region, based on near- and
mid-IR imaging and spectroscopy at a resolution of $\sim$\,0\farcs4
with the Subaru telescope. More recently, \citet{Hofmann04} studied
the morphology of the central $1''\times1''$\ with a resolution of
$\sim$\,0\farcs1 using speckle imaging at the SAO 6\,m telescope in Russia.

\begin{figure*}
\epsscale{2.1}
\plotone{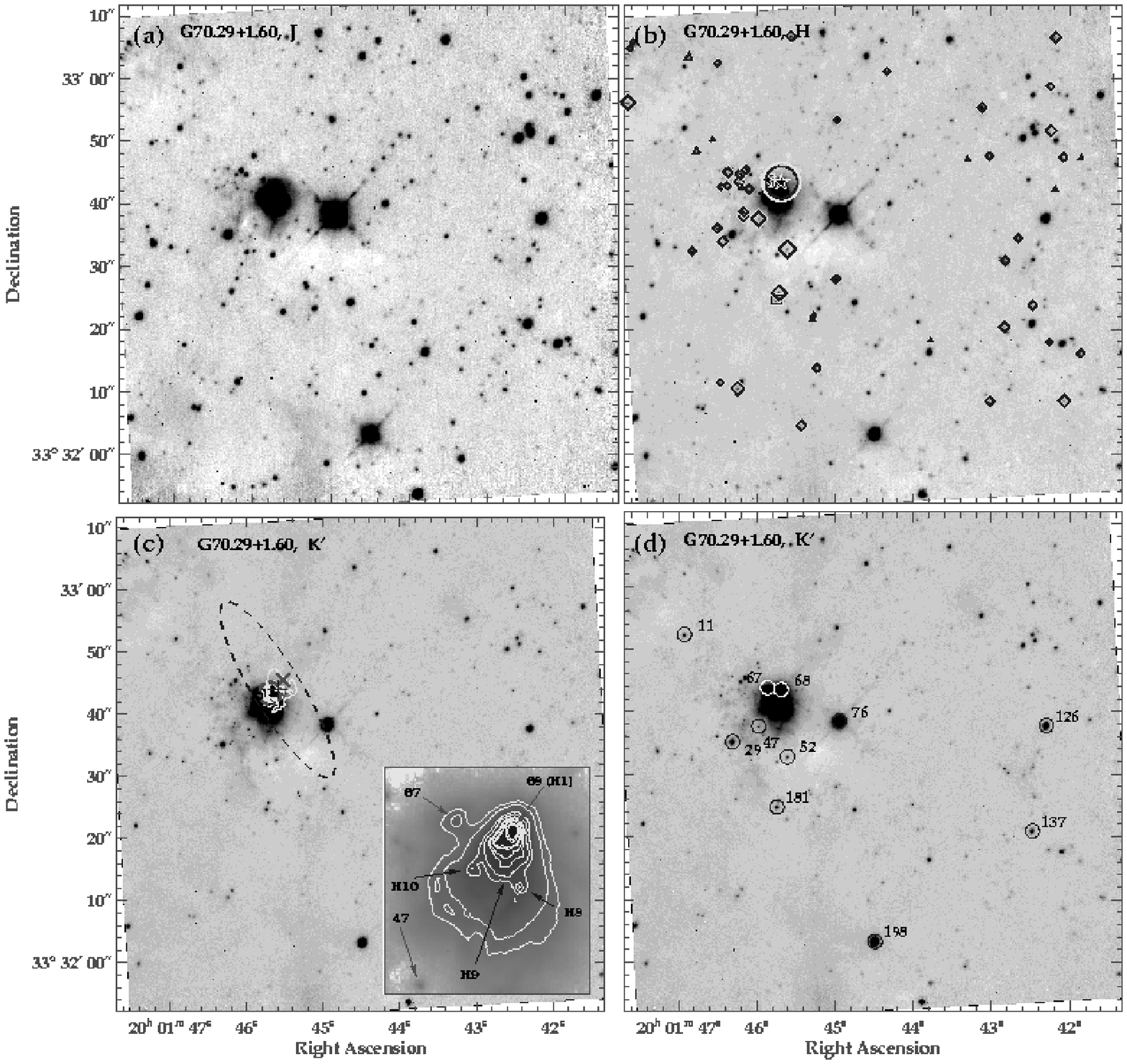}
\caption{Near-IR images of G70.29+1.60. 
	{\bf (a)} $J$-band image. The greyscale varies from 14.5 
	\mac\ (dark) to 14.9 \mac\ (white). {\bf (b)} $H$-band image. The
	greyscale varies from 13.2 to 13.8 \mac. See 
	Fig.~\ref{G309NearIRImLabel} for a key to the symbols.
	{\bf (c)} $K'$-image. The 
	greyscale varies from 10.6 to 10.8 \mac. The contours
	represent the radio-emission at 2 cm. Contour levels are at 
	5, 15 and 25$\sigma$. Symbols are the same as in
	Fig.~\ref{G309NearIRImLabel}. In this case, the MSX source,
	the centre of the IRAS ellipse, and the CS core coincide
	within the errors. The inset shows a close up of the
	central 7\farcs5$\times$8\farcs5 of G70.29+1.60. A 
	{\em loglog} scale was used to show both, the low level
	structure of the nebula as well as the point sources. Labels
	show some of the sources from \citet{Hofmann04} (H1, H8, H9,
	H10) and some of the stars studied in our photometry. Note
	that H1 and H10 correspond with sources OKYM3 and OKYM4 in
	\citet{Okamoto03}, respectively. 
	{\bf (d)} Some of the stars selected to 
	produce the colour-colour and colour-magnitude diagrams of 
	Figure~\ref{G70CCandCMLabel} are overlaid on the $K'$-band image. 
	\label{G70NearIRImLabel}}
\end{figure*}

\begin{figure*}
\epsscale{2.1}
\plotone{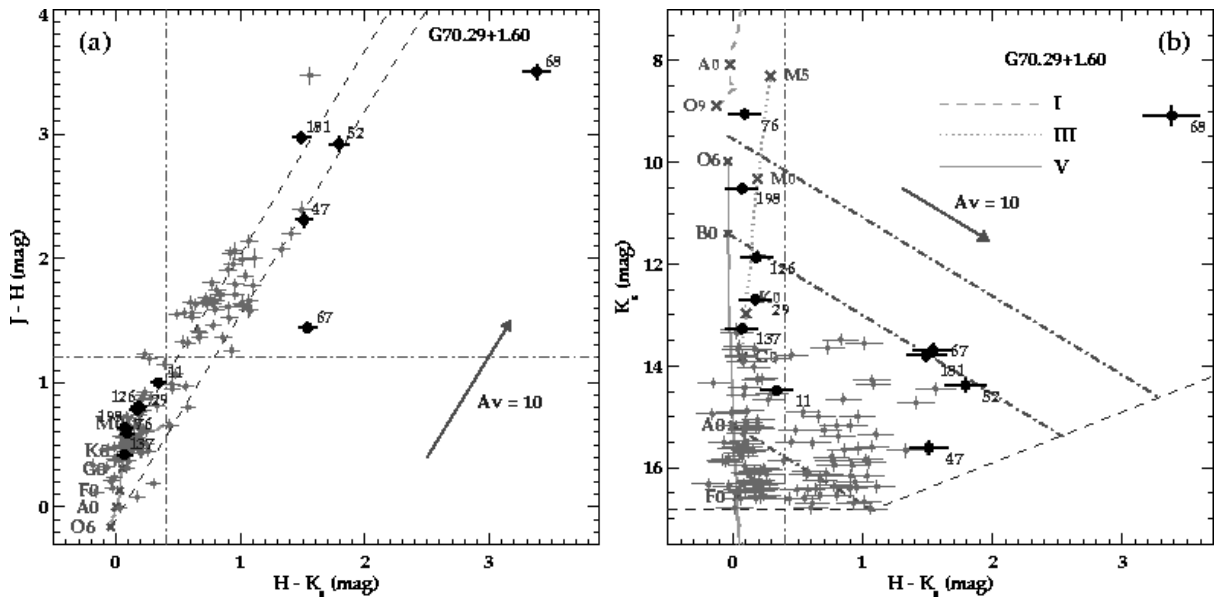}
\caption{{\bf (a)} Colour-colour and {\bf (b)} colour-magnitude 
	diagram of G70.29+1.60. Key to symbols and lines is given 
	in Figure~\ref{G309CCandCMLabel}.
	\label{G70CCandCMLabel}}
\end{figure*}

In Figure~\ref{G70NearIRImLabel}, we present our ALFA images of
\gsvo\ with a resolution of 0\farcs22 and a Strehl ratio
of 0.14 in the $K'$-band. The morphology of the near-IR nebula is 
similar to that found by \citet{Okamoto03} and \citet{Hofmann04} (see 
the inset in Fig.~\ref{G70NearIRImLabel}c).  Some of the sources
labelled in Figure~\ref{G70NearIRImLabel}d (\#29, \#47, \#52, \#67 and 
\#68) are clearly cross-identified with sources in Fig.~3 in 
\citet{Okamoto03} and Figs.~1 and~2 in \citet{Hofmann04}. 
We also encounter near-IR point sources coincident with the position of
the 11.4~\micron\ peaks OKYM\,3 and OKYM\,4. In
particular, our source \#68 is coincident with OKYM\,3, which is located
at the centre of the radio-emitting source. No photometry of
the near-IR source associated with OKYM\,4 was possible, due to its
location, which is highly embedded in the bright near-IR nebulosity.

Sources \#67 and \#68 are the most likely ionizing sources,
based on their location with respect to the radio peak and 
their position in the C-C and C-M diagrams. Source \#67 is consistent
with a B0V star under a visual extinction of $\sim$\,20 mag. Detailed
inspection of the $K'$~band image shows that this source is elongated in
the NW-SE direction. 
Source~\#68 appears to be extremely over-luminous in the C-M
diagram. It has an excess of $\sim$\,1.5~mag in the \hks\ colour. 
Even if we assume that this excess is mainly in the $K_s$\
band, the high luminosity of source \#68  can only be explained if
it is a super-giant of spectral type later than A0I under
$\sim$\,30~mag of visual extinction. However, due to the strong nebular
contamination within the region of 2\arcsec\ around \#68, this spectral
type estimate should be taken with caution. Besides, a star with such
a spectral type would not contribute to the ionization of the \hii\ region. 
Further high-resolution spectroscopy 
in the near-IR is required to find photospheric lines
associated with star \#68, and therefore, to determine whether this 
source is actually a star, and if this where the case, to determine 
its spectral type. 

In any case, we can compare the expected stellar ionizing population from
published radio and IR data with the one suggested by our near-IR
photometry. \citet{Kurtz94} estimated a log(\nlyc)=49.3 and 
\citet{MartinHernandez02} obtained a log(\nlyc)=49.1. If we scale these 
values to a distance of 8.2~kpc, we obtain an O5V spectral type for
a single ionizing star \citep{Smith02}. This value, based on 2~cm
interferometry, is in good agreement with the O6V - O9V spectral type 
inferred by \citet{Okamoto03} from modelling of the ionization
structure of the mid-IR source OKYM\,3 (i.e. our source
\#68). The Lyman photon rate of \gsvo\ (log(\nlyc)\,=\,49.2) can
also be produced by an O3I star, which would be in somehow better 
agreement with the luminosity inferred from our near-IR photometry. 
Regarding the other mid-IR source with near-IR counterpart (OKYM\,4), 
\citet{Okamoto03} obtained a spectral type between B0V and O9V. This 
source (star~10 in \citealp{Hofmann04}) is not included in our C-C 
and C-M diagrams because it was barely detected in the $K'$~image and 
not detected at all at shorter wavelengths. 

Regarding the total luminosity of \gsvo, its IRAS fluxes
indicate a $\mathrm{\log(L/L_{\odot})}$ $\sim6.30$. This cannot be
delivered by a single dwarf star
(note that $\mathrm{\log(L/L_\odot)}$\,=\,6.03 for an O3V star), but it 
is plausible for a single super-giant star
($\mathrm{\log(L/L_\odot)}$\,=\,6.27 for an O3I). Even source 
\#68, which maybe an A9I star, would be not luminous enough. This 
maybe due to the presence of a population of hot stars, hidden at 
near-IR wavelengths, which would be contributing to the 
IRAS emission. Some members of this population could correspond to 
the mid-IR sources detected by \citet{Okamoto03}.   

\clearpage

\clearpage

\subsubsection[G77.96-0.01]{\gsvsv}
\label{gsvsvSubSubSec}

This \uchii\ region, located at a distance of 4.2~kpc from the Sun,
was classified as irregular based on its morphology at radio
wavelengths \citep{Kurtz94}. Its sub-arcsecond morphology in the 
near-IR has not been previously studied.

\begin{figure*}
\epsscale{2.1}
\plotone{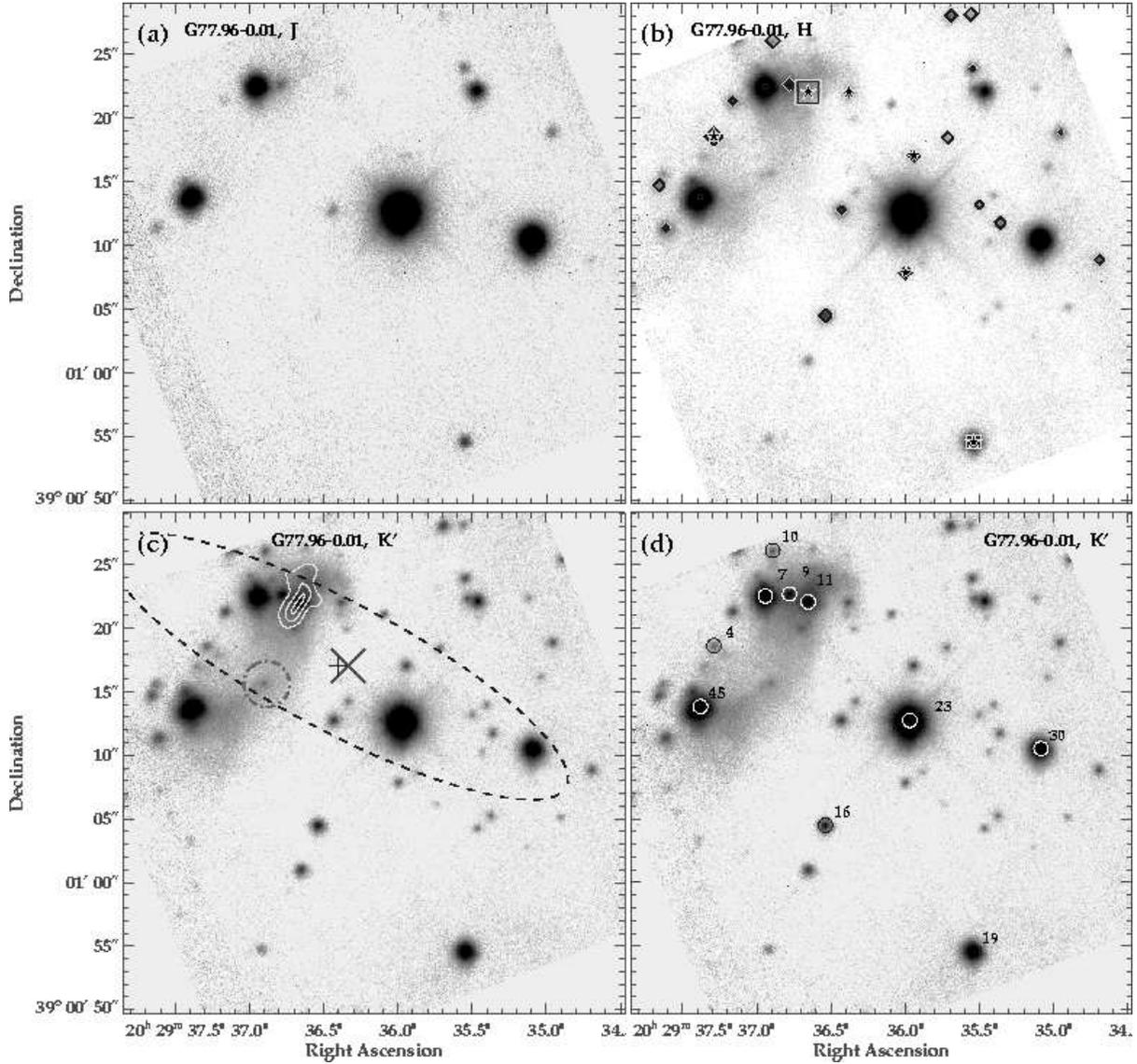}
\caption{Near-IR images of G77.96-0.01 with a resolution of $0\farcs5$. 
	{\bf (a)} $J$-band image. The greyscale varies from 13.4 
	\mac\ (dark) to 17.4 \mac\ (white). {\bf (b)} $H$-band image. The
	greyscale varies from 12.8 to 17.4 \mac. {See 
	Fig.~\ref{G309NearIRImLabel} for a key to the symbols.
	\bf (c)} $K'$-image. The 
	greyscale varies from 11.9 to 16.8 \mac. The contours
	represent the radio-emission at 3.6 cm. Contour levels are at 
	2, 6 and 9$\sigma$. Symbols are the same as in
	Fig.~\ref{G309NearIRImLabel}. {\bf (d)} Some of the stars selected to 
	produce the colour-colour and colour-magnitude diagrams of 
	Figure~\ref{G77CCandCMLabel} are overlaid on the $K'$-band image. 
	\label{G77NearIRImLabel}}
\end{figure*}

In Figure~\ref{G77NearIRImLabel}, we present the ALFA $J$, $H$\ and $K'$\
images of this region. The radio \uchii\ region is coincident with a
bright near-IR nebulosity in the north-eastern region of our
FOV. Inspection of the 2MASS $K_s$\ image over a field of view 5
times larger than the one shown in Figure~\ref{G77NearIRImLabel}
illustrates how the southern part of this reflection nebula is actually 
a bright knot of a long {\em snake-like} nebula extending for 
$\sim\,3'$\ towards the north-east and west. 

Our C-C and C-M diagrams indicate two likely candidates
which maybe ionizing the \hii\ region, sources \#7 and \#9, 
which are situated within 5\arcsec\ (i.e. 0.1~pc) of the \uchii\ 
region and within the near-IR nebulosity. They
have spectral types $\sim$\,O8V and B2V, respectively, both showing 
an extinction of about 10 magnitudes in the visual. At the location of
the radio-peak, we find source \#11. Its extinction is
high enough to yield only an upper limit for the J~magnitude. Its 
\hks\ colour shows an excess of $\sim$\,1~mag. Its spectral type after 
removing, this excess is that of an early BV star. Another possibility
is that this source may simply be an unresolved externally-ionized
knot. Source \#45 is also located within the near-IR nebulosity. The 
spectral type inferred for this source is O8V, with a visual
extinction of 10 mag. Therefore, two O8V stars and two early BV stars 
appear to be the the best candidates for ionizing sources of 
this region.

We estimate a log(\nlyc)\,=\,46.5 from the integrated
flux and size of the \gsvsv\ radio source at 2~cm \citep{Kurtz94}. This
value implies a B1V spectral type for the ionizing source, which is
later than the spectral types inferred from the near-IR photometry. For
the IRAS source associated with \gsvsv, we find a spectral type O8V, 
which is in good agreement with that of the hottest star in the
near-IR images. 


\clearpage

\begin{figure*}
\epsscale{2.1}
\plotone{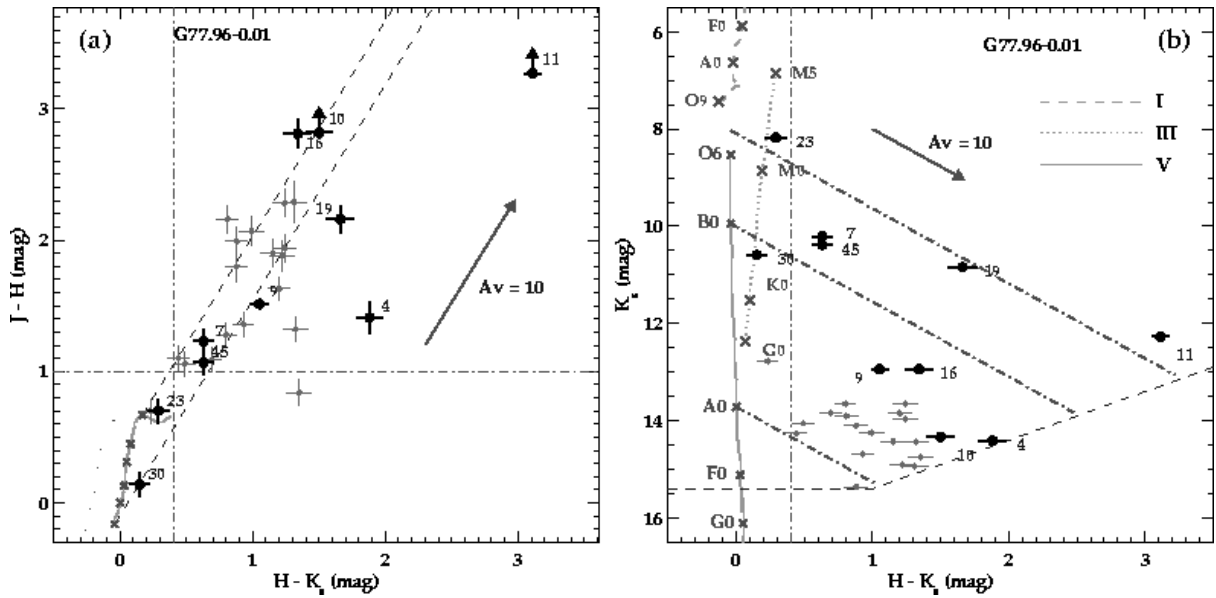}
\caption{{\bf (a)} Colour-colour and {\bf (b)} colour-magnitude 
	diagram of G77.96-0.01. Key to symbols and lines is given 
	in Fig.~\ref{G309CCandCMLabel}.
	\label{G77CCandCMLabel}}
\end{figure*}

\clearpage

\subsection[General Properties]{General Properties} 
\label{Results:Discussion}

After having discussed each source individually, in this section we
focus on the general features of our near-IR data, and compare them 
with published data at other wavelengths.

\subsubsection[Mass Function]{Mass Function}
\label{MassfunctionSec}

We note that all C-C diagrams show a gap in the distribution
of stars along the band that represents the extincted MS. This 
gap appears to be clear in \gsvo, but not so well defined
in \goe, where the photometric errors are larger. In any case, we
have used this gap as a criterion to differentiate between foreground
stars and stars which are likely to be associated with the \hii\
region. To quantify this distinction, we have defined two \jh\
and \hks\ colour cut-offs. Stars with colours above the cut-offs are
assumed to be at the same distance as the \uchii\ region. The spread of these
stars along the extinction band in the C-C diagram is considered to be
due to local variations of the dust and dense gas distribution at and 
in the surroundings of the \hii\ region. Colour cut-offs for each
region are listed in Table~\ref{StarCountsTable} and are shown as
dot-dashed lines in the C-C diagrams. In Table~\ref{StarCountsTable},
the {\em typical} extinction towards stars above the cut-off is listed 
for each region. This typical extinction was estimated from visual
inspection in the C-C diagram of the centroid of the stars along the 
reddened MS, which are also above the cut-offs. This value is slightly 
larger than the extinction associated to the point in the C-C diagram 
where both cut-off lines intersect. The expected line-of-sight
foreground extinction to each
source can be estimated from their distance. If we use 
\avi/d\,=\,0.68~mag\,kpc$^{-1}$\ from \citet{Bohlin78} (where d is the 
distance), the expected foreground extinction ranges
from 0.8~mag for \gtfo\ to 5.6~mag for \gsvo. In all \uchii{}s, this
foreground extinction is consistent with (i.e. smaller than) the \avi\ 
inferred from the colour cut-offs in our C-C diagrams.  

From the C-M diagrams, we counted the number of stars for each spectral
type located above the colour cut-offs, which we assume to be at the 
same distance as the \uchii\ region. The star counts are shown in
columns (4) to (8) of Table~\ref{StarCountsTable}. For the lower end
of the mass function, the completeness is clearly not reached, since
MS stars later than A are
normally below the detection limit. However, we can compare the ratio
of the number of O-type stars to the number of B-type stars (\novbv),
with the expected ratio from a theoretical expression for the IMF, 
since the completeness should be better in the upper end of the mass
spectrum. In this mass range, the exponent of the IMF is -1.3
\citep{Miller79}, yielding a ratio \novbv$\sim 0.16$. For the 
regions that have high stellar counts, i.e. \goe, \gso\ and 
\gsvo, the \novbv\ is 0.11, 0.03 and 0.07 respectively. The agreement
between these values and the IMF is reasonably good for \goe, but
fails for \gso\ and \gsvo. This maybe an indication that not all
\uchii\ regions contain very massive O stars. However, before any 
robust conclusion on the IMF can be drawn, deeper observations at 
higher resolution and also at longer wavelengths are needed to achieve 
a higher degree of completeness and to unveil very obscured stars. 

\clearpage
   \begin{deluxetable}{lccccccccc}
        \tabletypesize{\scriptsize}	
        \tablewidth{0pt}
	\tablecaption{Star Counts \label{StarCountsTable}}
	\tablehead{
	\colhead{Object} & \colhead{$(H - K_s)_c$ \tablenotemark{a}} &
	\colhead{$(J - H)_c$\,\tablenotemark{b}} & \colhead{$>$O3V\,\tablenotemark{c}} & \colhead{O3V-O9V}  &
	\colhead{B0V-B9V} & \colhead{A0V-A9V} & \colhead{$<$A9V\,\tablenotemark{c}} & \colhead{Ex.\,\tablenotemark{d}} & \colhead{\avi\,\tablenotemark{e}}\\
	\colhead{} & \colhead{mag} &
	\colhead{mag}  & \colhead{\#} & \colhead{\#}  & \colhead{\#} &
	\colhead{\#} & \colhead{\#} & \colhead{\#} & \colhead{mag}
	} 
	\startdata
            \gton      & 0.9  & 1.6  & 3  & 2  & 5   & \nodata & \nodata &  1 & 16 \\
            \gtfo      & 1.1  & 1.8  & 0  & 1  & 4   &    2    &    3    &  3 & 19 \\
            \gf        & 0.6  & 1.3  & 2  & 2  & 14  &    6    &    3    &  7 & 17 \\
            \goo       & 0.6  & 1.3  & 2  & 2  & 7   &    0    & \nodata &  3 & 16 \\
            \goe       & 0.6  & 1.4  & 0  & 3  & 27  &    2    & \nodata &  0 & 12 \\
            \gso       & 0.5  & 0.9  & 1  & 0  & 30  &   20    &    2    & 10 & 15 \\
            \gsvo      & 0.4  & 1.2  & 1  & 2  & 41  &   10    & \nodata &  2 & 14 \\
            \gsvsv     & 0.4  & 1.0  & 0  & 4  & 20  &    0    & \nodata &  6 & 14 \\
	\enddata
	\tablenotetext{a}{Cutoff in the $H-K_s$\ colour.} 
	\tablenotetext{b}{Cutoff in the $J-H$\ colour.} 
	\tablenotetext{c}{$>$O3V means earlier than O3V and $<$A9V
            means later than A9V.} 
	\tablenotetext{d}{Number of stars from columns (4) to (8) that 
	  show intrinsic excess in the C-C diagram.}
	\tablenotetext{e}{Typical visual extinction(s) towards 
	  stars with colours above the cut-offs inferred
          from the C-C diagrams. The \avi\ value for each region is
          calculated for a {\em typical} star above the cut-off, which 
	  is slightly larger than the \avi\ value for the cut-off itself.}
   \end{deluxetable}

\clearpage

\subsubsection[Near-IR Ionizing Sources]{Near-IR Ionizing Sources}
\label{IonizingSourcesSec}

We have made an attempt to identify, using near-IR photometry, the
main ionizing stars for each 
\uchii\ region. Column~(3) of Table~\ref{PhysicalParametersTable} shows that
for all regions we have found at least one source with a spectral type
early enough to be capable of ionizing an \hii\ region. However, the
question of whether this source(s) is the actual ionizing star(s) of the
radio \uchii\ region is still to be answered. Our
identification has the advantage of the high resolution delivered by
AO imaging and the relative transparency of dust and dense gas at near-IR
wavelengths. It is certainly true that these sources identified as 
ionizing stars are unresolved and show high extinction (\avi$>$30 mag) 
and/or intrinsic IR excess, which indicate their youth and possible relation
with the \hii\ region.  In all cases but \gtfo\ and \gso, their position 
coincides with that of the radio peak which defines the \uchii\
region (see H-band images in Figs.~\ref{G309NearIRImLabel} to 
\ref{G77NearIRImLabel}). Even the two exceptions can be
explained. \gtfo\ appears to be simply a knot which is part of a much larger
ionization shell ($1'$~radius, i.e 0.4~pc), so it is not surprising 
that no stars are found at the radio peak. \gso\ appears to be a
rather evolved compact \hii\ region, which shows several separated 
peaks at high-interferometric resolution and extends for over 30$''$\
(i.e. 0.4~pc) in low-resolution radio maps. 

We have used C-C and C-M diagrams to spectral-classify the possible 
ionizing sources and other sources associated with the \uchii\
regions. Our photometric spectral classification, however, is not without
problems. \uchii\ regions appear to be rather complex in the near-IR 
at sub-arcsecond scales, showing some degree of crowdness as 
well as spatially variable background due to the presence of extended 
nebulosity. This makes it difficult to calculate magnitudes accurately. 
To solve at least partially these problems, we have used PSF 
fitting photometry, which should reduce the contribution to the
stellar flux due to barely resolved stellar neighbours and due to
the presence of a spatially variable background. The possibility that 
some of the sources identified as single stars are multiple systems or 
unresolved emission line knots cannot be completely discarded due to 
the relatively large distances (between 1.2 and 8.2~kpc) to these 
\uchii\ regions. The high resolution delivered by AO though (between 
0\farcs13 and 0\farcs55, which turns into projected separations of 
$\sim\,1000$~AU), alleviates this possibility. Other typical problems 
associated with the photometric spectral classification are temporal 
variability in young stars, undetermined photometric excesses due to 
unresolved dust emission, scattering, accretion discs, and 
uncertainties in the distance and in the extinction.

\subsubsection[Comparison with Radio Sources]{Comparison with Radio Sources}
\label{ComparisonWithRadioSec}

In most of the \uchii{}s, the spectral type of the ionizing
source obtained from radio-continuum interferometry is a few sub-types
later than the spectral type inferred from our near-IR photometry. One 
possible explanation is that part of the ionizing flux produced by 
the sources detected in our near-IR images is being absorbed 
by dust in the region. If the radio free-free emission is optically thick, 
radio interferometry may also be underestimating the Lyman
continuum. A clumpy density distribution within the \uchii\ region can
also yield an underestimate for the \nlyc. One additional problem is the
fact that radio interferometry is sensitive only to high spatial
frequencies. If the \uchii\ region is a mere density enhancement
within a larger undetected extended halo of ionized gas, the total 
\nlyc\ associated with the full-size \hii\ region may be more compatible with 
the stellar content inferred from our near-IR photometry. Extended
halos of ionized gas are known to exist in a number of \uchii{}s (e.g. 
\citealt{Kurtz99, KimKoo01}). \citet{KimKoo01} find 
typical ratios $\ga 10$\ between H~{\sc i}\ 21~cm single dish fluxes and VLA
fluxes. Even though the higher continuum flux in the single-dish observations
comes from a larger ionized gas volume, the spectral types inferred
from single-dish observations are 4 times earlier than those inferred 
from radio-interferometry. Such kind of differences would help to reconcile 
the spectral types inferred from the near-IR stellar population with 
those inferred from radio measurements. 

A spectral type classification based on radio data also has the problem 
that it depends on the stellar models adopted. 
Table~\ref{PhysicalParametersTable} shows that for a given \nlyc, the 
spectral types inferred using the models in \citet{Smith02} are
earlier than the spectral types inferred from the models in 
\citet{Vacca96}. This is mainly due to the more realistic approach 
of \citet{Smith02}, by using models that include stellar winds,
spherical geometry, and line blanketing by metals
\citep{Pauldrach01}. Even though the line 
blanketing does not affect the \nlyc\,--\,\teff\ relation 
\citep[see ][]{Mokiem04}, it does affect the \teff\,--\,spectral 
type relation. This trend towards earlier spectral types for a given
\teff\ is also seen when the CMFGEN models \citep{Hillier98}, which
also include stellar winds and line blanketing, are used 
\citep[see ][]{Martins02}. Therefore, the use of models including a
more realistic physics helps to reconcile the near-IR photometric 
spectral-type classification of the possible ionizing stars and the
classification inferred from radio continuum measurements. 

\subsubsection[Comparison with IRAS Sources]{Comparison with IRAS Sources}
\label{ComparisonWithIRASSec}

We find that, in general, the luminosity inferred from the IRAS sources
associated with the \uchii{}s is lower than what it is expected from 
the analysis of the stellar population based on our near-IR photometry
(see Table~\ref{PhysicalParametersTable}). The spectral classification 
based on the luminosity inferred from IRAS fluxes is also model
dependent. In this case, Table~\ref{PhysicalParametersTable}
shows that for a given \ltot, the model grid by \citet{Smith02} yields 
earlier spectral types slightly earlier than the grid by \citet{Vacca96}.
This difference, even if it is only of half to one spectral sub-class, 
tends to reconcile the IRAS-based and photometric spectral types. 
In any case, even with the use of the grid by Smith et al., a 
discrepancy of over one spectral sub-class still exists between the
IRAS and near-IR spectral types for most of the sources.

One possible explanation for this general discrepancy between IRAS and 
near-IR luminosities is that the stellar population that shows up at 
near-IR wavelengths is already in a rather advanced evolutionary
stage, and therefore it is not the major contributor to the mid- and 
far-IR flux. It is likely that a more embedded population of stars
which would show up at longer wavelengths is actually producing the 
majority of the mid- and far-IR flux. This maybe the case in \gso\ and
\gsvo, where studies at wavelengths longer than 2\,\micron\ appear to
show part of this population (\citealp{Puga04a} and \citealp{Okamoto03}, 
respectively). AO-assisted observations with 8m-class telescopes at 
wavelengths 
longer than 2.2~\micron\  can provide the sensitivity and resolution 
necessary to pinpoint this possible embedded stellar
population \citep[see ][]{Feldt03}. 

Alternatively, if the dust spatial distribution is 
anisotropic, it is possible that the stars seen in the near-IR are actually
heating, but only partially, the dust producing the emission detected 
in the mid- and far-IR by IRAS. The radiation field associated with 
the IRAS source in this case will no longer be spherically symmetric, 
making luminosity estimates based on reprocessed radiation extremely 
geometry dependant. Stellar radiation will leak-out and leave no 
sign at mid- and far-IR wavelengths, yielding luminosities associated 
to the IRAS source smaller than those associated to the near-IR 
population. 

  A very simple calculation by dividing the luminosities inferred 
  from the IRAS fluxes and the luminosities associated with the
  spectral types inferred from our near-IR photometry 
  can be used to obtain a rough estimate of the fraction of 
  radiation that must be absorbed by the dust causing the IRAS 
  emission. We find that the fraction of absorbed radiation varies
  from only 10\% for \gtfo\ to as high as $\sim$\,85\% for \gton. It
  is $\sim$\,25\% for \gf, \goo\ and \gso B1, and between 55\% and
  65\% for \gsvsv\ and \goe. For \gsvo, the IRAS luminosity is higher
  than the near-IR luminosity. Interferometry in the mm and 
sub-mm range to trace the distribution of dust at a similar resolution 
to that of near-IR imaging, complemented with radiative transfer 
modelling will be necessary to further investigate this
calculation.

\subsubsection[CS Cores and MSX Sources]{CS Cores and MSX Sources}
\label{CSandMSXSec}

All \uchii{}s but \gtfo\ are associated
with a CS core listed in \citet{Bronfman96}. Their position
accuracy is $\sim\,3''$. \citet{Bronfman96} observed
these dense clumps in the millimetre CS(2-1) transition. They have sizes 
slightly higher than their instrumental beam size of $40'' - 50''$. The 
centroids of the CS cores are normally coincident with the IRAS
peaks. The typical separation between the near-IR ionizing
stars and the CS cores is between 7$''$\ and 24$''$ (0.1~pc to
0.45~pc, in terms of physical distances), with the exception of 
\gsvo, for which the separation is below 1$''$\ (i.e. below 8000~AU). 
If the detected near-IR 
massive stellar population is contributing to the heating and stirring
up (e.g. due to stellar winds) of the CS cores, one may expect
some correlation between the near-IR spectral type and the brightness
and velocity dispersion of the CS emission. We do not find such a 
correlation. The brightest CS core, with a mean
beam temperature (\tmb) of 5.6~K \citep{Bronfman96}, is associated
with \gf. However, this dense core is rather far away ($\sim$\,0.3~pc) 
from the main near-IR ionizing star (and also from the radio-peak of 
the \uchii\ region). In contrast, the near-IR ionizing star(s) of 
\gsvo, whose associated CS core has a relatively low \tmb\ (1.95~K), 
are coincident with the centroid of the CS emission. 
We note that the two cores with the broader CS(2-1) profiles are 
associated with \gf\ and \gsvo, which are known to be strong outflow 
sources.  


We also made an attempt to correlate the properties of the MSX
sources associated with the \uchii\ regions in our sample, and the
near-IR population. Inspection of our images shows that the centres of 
the MSX peaks are closer to the location of the radio \uchii\ regions
than the CS cores and IRAS sources. In column~(9) of
Table~\ref{PhysicalParametersTable}, we show the 
luminosity associated with the MSX sources. This luminosity was
calculated by integrating a black body fit using only the fluxes 
at 14.65~\micron\ and 21.34~\micron\ (see Sec.~\ref{Physics:IRASMSX}), 
respectively. We find no correlation between the luminosity of the 
MSX source and the luminosity of the near-IR ionizing population. 

\subsubsection[Possible Super-giants?]{Possible Super-giants?}
\label{SuperGiantsSec}

We note that in 3 of the regions (\gton, \gso{}B1 and \gsvo) the 
most luminous identified sources have near-IR colours consistent 
with being super-giants, rather than dwarfs. In the case of the
regions \gton\ and \gso{}B1, these
stars appear to have spectral types early enough as to be responsible 
for the ionization of the \hii\ region. A caveat should be pointed out 
here, since these high-luminosity sources may as well be mere unresolved 
ionized knots externally illuminated by other stars, rather than
actual stars. Note that for instance, an ionized knot emitting purely
in Br$\gamma$\ may appear as a strong continuum source in the broad-band
\ks\ filter. If they are actually stars, their large absolute
magnitude in K band (M$_K$) would be likely to be generated by a
single star, since it cannot be explained as due to an 
unresolved multiple system formed by two or more stars. To prove this, 
we analyse the most exteme case, i.e. two O3V stars being
misinterpreted as an O9.5I star. The visual magnitude (M$_v$)\ 
and intrinsic colours for an O9.5I star and an 03V star are such that 
M$_K$(O9.5I)\,=-5.65 mag and M$_K$(O3V)\,=-4.79 mag, respectively 
\citep{Vacca96,Aller82}. An unresolved binary composed of two O3V
stars would have a total M$_K$=-5.24 mag, which
is still 0.4~mag fainter than an O9.5I star. This difference is larger than
the typical error in our photometry. Nevertheless, this argument cannot be
used to discard possible triple and multiple systems of dwarfs.
In one particular case (\gso) further Br$\gamma$\
and $L'$\ imaging strongly supports the interpretation of the most
luminous source being a super-giant rather than a dwarf 
\citep{Puga04a}. 

Although further near-IR spectroscopy is required to confirm the
nature of these high-luminosity sources \citep{Kaper02}, we can
argue whether, in terms of evolution, the presence 
of a super-giant in these \hii\ regions is plausible. We use Eq.~(9) from
\citet{wc89}, with a  typical value for the initial radius of the 
Str\"{o}mgren sphere of 0.02~pc and a typical sound speed of 
10~\kms. If we take into account the physical radius of 
\gso{}B1, \gsvo\ and \gton, i.e. 0.18~pc, 0.14~pc and 0.03~pc
respectively, then the ages of these \hii\ regions are $5\times10^4$,
$2.5\times10^4$\ and $1.1\times10^3$\ years, respectively. 
This is between 2 and 3 orders of magnitude shorter than the typical 
time of $\la 10^6$~yr required for a 40~\msun\ star to reach the 
super-giant phase \citep{Schaerer97}. 

Several arguments can be used to re-concile the presence of
super-giants in some of the \uchii\ regions. One may
think that these luminous stars do not belong to the \uchii\ region,
but that they are foreground stars situated at the position of the radio
peak by chance allignments. This is rather unlikely though, since they 
tend to be the most embedded sources in our C-C diagram. Besides, even 
though our sample is rather small, we find possible ionizing
super-giants in 25\% 
of the cases, which is difficult to explain simply by chance 
alignments. The presence of super-giants could be explained if the 
\hii\ regions were actually in a more advanced evolutionary stage than 
what is inferred from their physical size. This maybe the case if radio 
single-beam observations (or low resolution interferometry) of the ionized 
gas showed the presence of extended halos associated with these \uchii\ 
regions. However, in instances as \gso, even with a low-resolution
configuration of the VLA, the 3.6~cm emission extends only for
0.18~pc. The external pressure due to the molecular cloud in which
\uchii{}s are normally embedded may delay the expansion of the
ionization front, yielding also older ages than what is predicted by 
simple expansion models. This argument has already been used before to
explain the paradigm of the \uchii{}s lifetime \citep{wc89}.
\citet{Tan04} show that an \hii\ region will expand to a given 
  size in a timescale 2 orders of magnitude slower in a clumpy
  turbulent medium than in a uniform medium.

\section[CONCLUSIONS]{CONCLUSIONS}
\label{Conclusions}

We have analized AO-assisted $J$, $H$\ and $K/K'$\ band images and
photometry of 8 \uchii\ regions. These data show details of the
near-IR morphology and stellar population in these regions with 
unprecedented spatial resolution. We have analized the stellar
population using near-IR photometry. 
We find that the typical spectral types obtained from near-IR
photometry are earlier than those predicted from radio-continuum 
interferometric data available in the literature. The possible 
underestimate of the rate of Lyman photons from radio data due 
to absorption by dust, optical depth effects, clumpiness or the
presence of undetected extended ionized halos may be causing this 
difference. We have also compared the ionizing and luminous 
near-IR stellar population with the luminosity of the IRAS source 
associated with each \uchii\ region. In general, we find that IRAS 
fluxes yield later spectral types for the dust heating source 
than near-IR photometry. This maybe due to the presence of a more 
embedded stellar population than the one accessible with near-IR 
photometry. This hidden population would be the main contributor to 
the IRAS luminosity. Alternatively, if the dust distribution is 
anisotropic, the stars seen in the near-IR may actually be heating 
the dust seen by IRAS. Stellar models with spherical geometry and 
a detailed treatment of stellar winds and line blanketing 
yield radio- and IRAS-based spectral types that are generally 
in better agreement with photometry-based spectral types than 
non-LTE, plane-parallel models.

In all cases, the C-C diagrams for these regions show two
groups of stars. We assume that one of them, close to the unreddened 
theoretical Main Sequence, is formed by foreground stars. The stars 
in the other one normally show a large spread in extinction, and 
they are certainly associated with the \uchii\ region. In all regions, we
find at least one source with a spectral type early enough to be the
ionizing star of the \hii\ region, i.e. early B/O. We have counted, in 
spectral-type bins, all stars associated with each \uchii\ region. For 
\goe, where we have a relatively high nuber of stars, a reasonable
agreement is found between the observed O and B star numbers and those 
predicted by the IMF given in \citet{Miller79}. However, this is not 
the case for \gso\ and \gsvo. For the remaining \uchii{}s, not 
enough stellar counts were available to make any estimate.


We have also compared the luminosity of the near-IR ionizing stellar 
population with MSX luminosities and no correlation was found. 
Our near-IR data were compared with available millimetre CS data
to find out what is the impact the massive stellar population in 
\uchii{}s has on the surrounding dense gas. No correlation was found 
between the brightness of the CS emission and the luminosity of 
the best near-IR candidates to ionize the \uchii\ region. We note 
that the velocity dispersion is larger in the CS cores in the \gf\ 
and \gsvsv\ regions, which are known to be associated with massive 
outflows. 

In 3 regions (\gso{}B1, \gsvo\ and \gton), the most luminous stars  
in the \uchii\ region appear to be super-giants. In \gso{}B1 and
\gton, these luminous stars are also the best candidates for 
ionizing sources of the \hii\ region. 

The main conclusion from this work is that AO-assisted near-IR imaging
of \uchii{}s with 4m-class telescopes is a useful tool to study the 
stellar population in massive star-forming regions. This appears 
to be a more accurate method to study the massive young stellar population 
than any indirect method based on radio, IRAS or MSX measurements, mainly
due to the higher resolution provided by AO observations. Nevertheless,
it is also fundamental to observe at longer wavelengths 
(from 3.5~\micron\ to 20~\micron) with similar spatial resolution in order 
to detect the most embedded stellar population. Deeper observations 
are necessary to achieve a better coverage of the IMF. Finally, 
broad-band imaging should be complemented with near-IR spectroscopy 
and narrow-band imaging to fully understand the nature of the relevant
stellar population in each region. AO imaging at 8m-class telescopes 
is a promising tool to achieve these goals.



\acknowledgments
We thank the staff at the Calar Alto and La Silla observatories for 
their invaluable assistance during the observations. We also 
thank M. Kasper for his support during ALFA comissioning and
science verification. C. Alvarez would like to 
thank N. L. Mart\'{i}n-Hern\'{a}ndez for her careful reading of the 
manuscript and useful comments. We thank the
anonymous referee for the careful reading and useful comments, which have
helped to improve the quality of this manuscript. This research has 
made use of the NASA/IPAC Infrared Science Archive, which is operated 
by the Jet Propulsion Laboratory, California Institute of Technology, under
contract with the National Aeronautics and Space Administration. This 
research has made use of the SIMBAD database, operated at CDS,
Strasbourg, France.

\clearpage


\begin{thebibliography}{}

\bibitem[Acord et al.(1997)]{Acord97} {Acord}, J.~M., {Walmsley},
  C.~M., \& {Churchwell}, E. 1997, \apj, 475, 693
\bibitem[Afflerbach et al.(1996)]{Afflerbach96} {Afflerbach}, A., 
  {Churchwell}, E., {Acord}, J.~M., et al. 1996, \apjs, 106, 423
\bibitem[Aller et al.(1982)]{Aller82} {Aller}, L.~H., {Appenzeller}, 
  I., {Baschek}, B., et al. 1982, Landolt-B{\" o}rnstein: Numerical 
  Data and Functional Relationships in Science and Technology, p451 
\bibitem[Argon et al.(2000)]{Argon00} {Argon}, A.~L., {Reid},
  M.~J., \& {Menten}, K.~M. 2000, \apjs, 129, 159
\bibitem[Behrend \& Maeder(2001)]{behrend:2001} {Behrend}, R. \&
  {Maeder}, A. 2001, A\&A,  373, 190 
\bibitem[Beuzit et al.(1994)]{Beuzit94} {Beuzit}, J., {Hubin}, N.,
  {Gendron}, E., {Demailly}, L., et al. 1994, SPIE, 2201, 955
\bibitem[Blair et al.(1975)]{Blair75} {Blair}, G.N., {Peters} \& {van
  den Bout}, P. A. 1975, \apj, 200, L61 
\bibitem[Blitz et al.(1982)]{Blitz82} {Blitz}, L., {Fich}, M. \&
  {Stark}, A.~A. 1982, \apjs, 49, 183 
\bibitem[Bonnell et al.(1998)]{bonnell:98} Bonnell I.~A., Bate M.~R., 
  Zinnecker H. 1998, MNRAS,  298, 93 
\bibitem[Bohlin et al.(1978)]{Bohlin78} {Bohlin}, R.~C., {Savage},
  B.~D., \& {Drake}, J.~F. 1978, \apj, 224, 132
\bibitem[Braz \& Epchtein(1983)]{Braz83} {Braz}, M.~A. \&
  {Epchtein}, N. 1983, A\&AS, 54, 167
\bibitem[Bronfman et al.(1996)]{Bronfman96} {Bronfman}, L., {Nyman},	
  L.-A., \& {May}, J. 1996, 115, 81  
\bibitem[Burton et al.(2000)]{Burton00} {Burton}, M.~G., {Ashley},
  M.~C.~B., {Marks}, R.~D., et al. 2000, \apj, 542, 359
\bibitem[Cesaroni et al.(1991)]{Cesaroni91}  {Cesaroni}, R.,
  {Walmsley}, C.~M., \& {K\"ompe}, C., et al. 1991, \aap, 252, 278
\bibitem[Churchwell et al.(1978)]{Churchwell78} {Churchwell}, E., {Smith},
  L.~F., \& {Mathis}, J., et al. 1978, \aap, 70, 719
\bibitem[Crowther \& Dessart(1998)]{Crowther98a} {Crowther}, P.~A.
  \& {Dessart}, L. 1998, \mnras, 296, 622
\bibitem[Crowther(1998)]{Crowther98b} {Crowther}, P.~A. 1998, 
  The effective temperatures of hots stars, IAU Symp. 189: 
  Fundamental Stellar Properties, 137
\bibitem[Crowther \& Conti(2003)]{Crowther03}  {Crowther}, P. \&
  {Conti}, P.~S. 2003, \mnras, 343, 143
\bibitem[Cutri et al.(2003)]{Cutri03} {Cutri}, R.~M.,
  {Skrutskie}, M.~F., {van Dyk}, S., {Beichman}, C.~A., {Carpenter}, J.~M.
  et al. 2003, 2MASS All-Sky Catalog of Point Sources, VizieR, II/246 
\bibitem[Deharveng et al.(2000)]{Deharveng00} {Deharveng}, L.,
  {Nadeau}, D., {Zavagno}, A., et al. 2000, \aap, 360, 1107
\bibitem[DePree et al.(1994)]{dePree94} {DePree}, C. G., {Goss},
  W. M., {Palmer}, P., et al. 1984, \apj, 428, 670 
\bibitem[Ducati et al.(2001)]{Ducati01} {Ducati}, J.~R., {Bevilacqua}, 
  C.~M., {Rembold}, S.~B., \& {Ribeiro}, D. 2001, \apj, 558, 309
\bibitem[Egan et al.(2003)]{Egan03} {Egan}, M.~P., {Price}, S.~D., 
{Kraemer}, K.~E., \& {Mizuno}, D.~R., et al 2003, MSX6C Infrared Point 
Source Catalog, VizieR, V/114 
\bibitem[Epchtein \& Lepine(1981)]{EpchteinLepine81} {Epchtein}, N.
  \& {Lepine}, J.~R.~D. 1981, \aap, 99, 210 
\bibitem[Epchtein et al.(1981)]{Epchtein81} {Epchtein}, N.,
  {Guibert}, J., {Rieu}, N.~Q., et al. 1981, \aap, 97, 1 
\bibitem[Evans et al.(1981)]{Evans81} {Evans}, N. J., {Harvey}, P.,
  {Israel}, F., et al. 1981, \apj, 250, 200
\bibitem[Feldt et al.(1998)]{feldt:98} Feldt M., Stecklum B., Henning 
  Th., Hayward T.~L., Lehmann T., \& Klein R. 1998, A\&A,  339, 759 
\bibitem[Feldt et al.(1999)]{feldt:99} {Feldt}, M., {Stecklum}, B.,
  {Henning}, Th., et al. 1999, \aap, 346, 243
\bibitem[Feldt et al.(2003)]{Feldt03} {Feldt}, M., {Puga}, E.,
  {Lenzen}, R., {Henning}, Th., et al. 2003, \apjl, 599, L91
\bibitem[Felli \& Harten(1981)]{Felli81} {Felli}, M. \& {Harten},
  R. H. 1981, \aap, 100, 42
\bibitem[White \& Fridlund(1992)]{WhiteFridlund92} {White}, G. J. 
  \& {Fridlund}, C. V. M. 1992, \aap, 266, 252
\bibitem[Garay et al.(1993)]{Garay93} {Garay}, G., {Rodriguez}, L. F., 
  {Moran}, J. M., \& {Churchwell} E. 1993, \apj, 418, 368 
\bibitem[Garay et al.(1994)]{Garay94} {Garay}, G., {Lizano}, S., 
  \& {Gomez}, Y. 1994, \apj, 429, 268 
\bibitem[Garay et al.(1998a)]{Garay98a} {Garay}, G., {Gomez},
  Y., {Lizano}, S., et al. 1998a, \apj, 501, 699 
\bibitem[Garay et al.(1998b)]{Garay98b} {Garay}, G., {Gomez},
  Y., {Lizano}, S., et al. 1998b, \apj, 501, 710 
\bibitem[Hanson et al.(2002)]{Hanson02} {Hanson}, M.~M., {Luhman},
  K.~L., \& {Rieke}, G.~H. 2002, \apjs, 138, 35
\bibitem[Harvey \& Gatley(1983)]{HarveyGatley83} {Harvey}, P.~M.
  \& {Gatley}, I. 1983, \apj, 269, 613
\bibitem[Harvey \& Wilking(1984)]{HarveyWilking84} {Harvey}, P.~M.
  \& {Wilking}, B.~A. 1984, \apjl, 280, L19
\bibitem[Harvey \& Forveille(1988)]{Harvey88} {Harvey}, P.\ M. \&
  {Forveille}, T. 1988, \aap, 197, L19
\bibitem[Harvey et al.(1994)]{Harvey94} {Harvey}, P.~M., {Lester},
  D.~F., {Colome}, C., et al. 1994, \apj, 433, 187
\bibitem[Henning et al.(1990)]{Henning90} {Henning} Th., {Pfau}, W., 
  \& {Altenhoff}, W. J. 1990, \aap, 227, 542 
\bibitem[Henning et al.(2001)]{henning:2001} Henning Th., Feldt M., 
  Stecklum B., \& Klein R. 2001, \aap,  370, 100 
\bibitem[Henning et al.(2002)]{Henning02} Henning, Th., Stecklum, B.,
  \& Feldt, M. 2002, ASP Conf.Ser. 267, 153 
\bibitem[Hillier \& Miller(1998)]{Hillier98} {Hillier}, D.~J. \& {Miller},
  D.~L. 1998, \apj, 496, 407
\bibitem[Hippler et al.(1998)]{hippler:98} Hippler S., Glindemann A.,
  \& Kasper M. 1998, SPIE Proc.\ 3353, 44
\bibitem[Hofner \& Churchwell(1996)]{Hofner96} {Hofner}, P. \&
  {Churchwell}, E. 1996, \aaps, 120, 283
\bibitem[Hofmann et al.(1995)]{hoffmann:95} Hofmann, R., Brandl, B.,
  Eckart, A., Eisenhauer, F., \& Tacconi-Garman, L.E. 1995, SPIE, 2475,
  192
\bibitem[Hofmann et al.(2004)]{Hofmann04} {Hofmann}, K., {Balega},
  Y., {Preibisch}, T., \& {Weigelt}, G. 2004, \aap, 417, 981
\bibitem[Howard et al.(1996)]{Howard96} {Howard}, E. M., {Pipher},
  J. L., {Forrest}, W. J., et al. 1996, \apj, 460, 744
\bibitem[Jackson \& Kraemer(1999)]{Jackson99} {Jackson}, J.\ M. \& 
  {Kraemer}, K.\ E. 1999, \apj, 512, 260
\bibitem[Kaper et al.(2002)]{Kaper02} {Kaper}, L., {Bik}, A.,
 {Hanson}, M.~M., \& {Comer{\' o}n}, F. 2002, ASP Conf. Ser., 267, 95
\bibitem[Kim \& Koo(2001)]{KimKoo01} {Kim}, K. \& {Koo}, B. 2001, \apj,
  549, 979 
\bibitem[Kurtz et al.(1994)]{Kurtz94} {Kurtz}, S., {Churchwell}, E., \&
  {Wood}, D.~O.~S. 1994, \apjs, 91, 659
\bibitem[Kurtz et al.(1999)]{Kurtz99} {Kurtz}, S.~E., {Watson}, A.~M., 
  {Hofner}, P., et al. 1999, \apj, 514, 232
\bibitem[Lenzen et al.(1998)]{lenzen:98} Lenzen, R., Bizenberger, P., Salm, 
  N., \& Storz, C. 1998, SPIE,  3354, 493 
\bibitem[Lockman (1989)]{Lockman89} {Lockman}, F.~J. 1989, \apjs, 71, 469 
\bibitem[Loughran et al.(1986)]{Loughran86} {Loughran}, L.,
  {McBreen}, B., {Fazio}, G.~G., et al. 1989, \apj, 303, 629
\bibitem[Lumsden et al.(2002)]{Lumsden02} {Lumsden}, S.~L., {Hoare},
  M.~G., {Oudmaijer}, R.~D., et al. 2002, \mnras, 336, 621 
\bibitem[Martin-Hernandez et al.(2002)]{MartinHernandez02} 
  {Mart{\' i}n-Hern{\' a}ndez}, N.~L., {Peeters}, E., {Morisset},
  C., et al. 2001, \aap, 381, 606
\bibitem[Martins et al.(2002)]{Martins02} {Martins}, F., {Schaerer},
  D., \& {Hillier}, D.~J. 2002, \aap, 382, 999 
\bibitem[McBreen et al.(1979)]{McBreen79} {McBreen}, B., {Fazio},
  G.~G., {Stier}, M., et al. 1979, \apjl, 232, 183
\bibitem[McKee \& Tan(2002)]{mckee:2002} McKee C.~F. \& Tan J.~C. 2002, 
  Natur,  416, 59 
\bibitem[Miller \& Scalo(1979)]{Miller79} {Miller}, G.~E. \&
  {Scalo}, J.~M. 1979, \apjs, 41, 513 
\bibitem[Mokiem et al.(2004)]{Mokiem04} {Mokiem}, M.R., 
  {Mart{\'{\i}}n-Hern{\' a}ndez}, N.L., {Lenorzer}, A., {de Koter},
  A., \& {Tielens}, A.G.G.M 2004, \aap, 419, 319  
\bibitem[Nakano et al.(2000)]{nakano:2000} Nakano T., Hasegawa T., Morino 
  J., \& Yamashita T. 2000, ApJ, 534, 976 
\bibitem[Okamoto et al.(2003)]{Okamoto03} {Okamoto}, Y., {Kataza},
  H., {Yamashita}, T., et al. 2003, \apj, 584, 368 
\bibitem[Panagia(1973)]{Panagia73} {Panagia}, N. 1973, \aj, 78, 929 
\bibitem[Pauldrach et al.(2001)]{Pauldrach01} {Pauldrach}, A.~W.~A., 
  {Hoffmann}, T.~L., \& {Lennon}, M. 2001, \aap, 375, 161 
\bibitem[Peeters et al.(2002)]{Peeters02} {Peeters}, E., 
  {Mart{\' i}n-Hern{\' a}ndez}, N.~L., {Damour}, F., et al. 2002, \aap, 
  381, 571
\bibitem[Phillips \& Mampaso(1991)]{Phillips91} {Phillips}, J.\ P. \& 
  {Mampaso}, A. 1991, \aaps, 88, 189
\bibitem[Puga et al.(2004a)]{Puga04a} {Puga}, E., {Alvarez}, C.,
  {Feldt}, M., {Henning}, Th., et al. 2004a, \aap, submitted
\bibitem[Puga et al.(2004b)]{Puga04b} {Puga}, E., et al., 
  2004b, in prep.
\bibitem[Rengarajan \& Ho(1996)]{Rengarajan96} {Rengarajan}, T.~N.
  \& {Ho}, P.~T.~P. 1996, \apj, 465, 363
\bibitem[Roelfsema et al.(1988)]{Roelfsema88} {Roelfsema}, P.~R.,
  {Goss}, W.~M., \& {Geballe}, T.~R. 1988, \aap, 207, 132
\bibitem[Rubin \& Turner(1969)]{RubinTurner69} {Rubin}, R. H. \&
  {Turner}, B. E. 1969, \apj, 157, 41
\bibitem[Simon et al.(1985)]{Simon85} {Simon}, T., {Dyck}, H.\ M., 
  {Wolstencroft}, R.\ D. et al. 1985, \mnras, 212, 21
\bibitem[Smith et al.(2002)]{Smith02} {Smith}, L.~J., {Norris},
  R.~P.~F., \& {Crowther}, P.~A. 2002, \mnras, 337, 1309
\bibitem[Schaerer \& de Koter(1997)]{Schaerer97} {Schaerer}, D. \& 
  {de Koter}, A. 1997, \aap, 322, 598
\bibitem[Schwartz et al.(1973)]{Schwartz73} {Schwartz}, P. R., {Wilson},
  W. J., \& Epstein, E. E. 1973, \apj, 186, 539 
\bibitem[Shepherd \& Churchwell(1996)]{Shepherd96} {Shepherd}, D.~S. 
  \& {Churchwell}, E. 1996, \apj, 457, 267
\bibitem[Sollins et al.(2004)]{Sollins04} {Sollins}, P. K., {Hunter}, T. R., 
  {Battat}, J., {Beuther}, H., et al. 2004., to be published in \apjl 
  (astro-ph/0403524)
\bibitem[Solomon et al.(1987)]{Solomon87} {Solomon}, P.~M., {Rivolo},
  A.~R., {Barrett}, J., et al. 1987, \apj, 319, 730
\bibitem[Straw et al.(1989)]{Straw89} {Straw}, S. M., {Hyland},
  A. R., \& {McGregor}, P. J. 1989, \apjs, 69, 99 
\bibitem[Tan \& McKee(2004)]{Tan04} {Tan}, J.~C. \& {McKee}, C.~F. 
  2004, Cancun Workshop on Formation and Evolution of Young Massive 
  Clusters, eds. H. Lamers, A. Nota and L. Smith, (astro-ph/0403498)
\bibitem[Tokunaga(2000)]{Tokunaga00} Tokunaga, A. T. 2000, Allen's
  Astrophysical Quantities, 4th edition, ed. A.~N. Cox, 143
\bibitem[Turner \& Matthews(1984)]{Turner84} {Turner}, B. E.
  \& {Matthews}, H. E. 1984, \apj, 277, 164
\bibitem[Vacca et al.(1996)]{Vacca96} {Vacca}, W.~D., {Garmany},
  C.~D., \& {Shull}, J.~M. 1996, \apj, 460, 914
\bibitem[Walsh et al.(1997)]{Walsh97} {Walsh}, A.~J., {Hyland},
  A.~R., \& {Robinson}, G. 1997, MNRAS, 291, 261
\bibitem[Walsh et al.(1998)]{Walsh98} {Walsh}, A.~J., {Burton}, M.~G.,
  {Hyland}, A.~R., et al. 1998, MNRAS, 301, 640 
\bibitem[Walsh et al.(1999)]{Walsh99} {Walsh}, A.~J., {Burton},
  M.~G., {Hyland}, A.~R., et al. 1999, MNRAS, 309, 905
\bibitem[Wink et al.(1983)]{Wink83} {Wink}, J. E., {Willson},
  T. L., \& {Bieging}, J. H. 1983, \aap, 127, 211 
\bibitem[Wolfire \& Cassinelli(1987)]{wolfire:1987} Wolfire M.~G. 
  \& Cassinelli J.~P. 1987, ApJ, 319, 850 
\bibitem[Wolstencroft et al.(1987)]{Wolstencroft87} {Wolstencroft},
  R.\ D., {Scarrott}, S.\ M., \& {Warren-Smith}, R.\ F. 1987, \mnras,
  228, 805 
\bibitem[Wood \& Churchwell(1989)]{wc89} Wood D.~O.~S. \& Churchwell E. 
  1989, ApJS, 69, 831 
\bibitem[Wouterloot \& Branz(1989)]{WouterlootBranz89} {Wouterloot},
  J.~G.~A. \& {Brand}, J. 1989, \aaps, 80, 149 
\bibitem[Wynn-Williams et al.(1977)]{WynnWilliams77} {Wynn-Williams},
  C. G., {Becklin}, E. E., {Matthews}, K., et al. 1977, \apj, 342, 860 
\bibitem[Yorke \& Sonnhalter(2002)]{yorke:2002} Yorke H.~W. \& 
 Sonnhalter C. 2002, ApJ, 569, 846 

\end{thebibliography}
\end{document}